\documentclass[lettersize,journal]{IEEEtran}
\usepackage{amsmath,amsfonts}
\usepackage{algorithmic}
\usepackage{algorithm}
\usepackage{array}
\usepackage[caption=false,font=normalsize,labelfont=sf,textfont=sf]{subfig}
\usepackage{textcomp}
\usepackage{stfloats}
\usepackage{url}
\usepackage{verbatim}
\usepackage{graphicx}
\usepackage{cite}
\usepackage{glossaries}
\usepackage{balance}
\usepackage{amssymb}
\usepackage{xcolor}
\newacronym{asv}{ASV}{autonomous surface vehicle}
\newacronym{auv}{AUV}{autonomous underwater vehicle}
\newacronym{uav}{UAV}{unmanned aerial vehicle}
\newacronym{ber}{BER}{bit-error-rate}
\newacronym{fov}{FoV}{field-of-view}
\newacronym{rf}{RF}{radio-frequency}
\newacronym{uwoc}{UWOC}{underwater wireless optical communication}
\newacronym{w2a}{W2A}{water-to-air}
\newacronym{rte}{RTE}{radiative transfer equation}
\newacronym{ook}{OOK}{on-off keying}
\newacronym{mimo}{MIMO}{multiple-input multiple-outputs}
\newacronym{fec}{FEC}{forward error correction}
\newacronym{mbps}{Mbps}{megabits per second}

\newacronym{ld}{LD}{laser diode}
\newacronym{led}{LED}{light-emitting diode}
\newacronym{apd}{APD}{avalanche photodiode}
\newacronym{mppc}{MPPC}{multi-pixel photon counter}
\newacronym{sipm}{SiPM}{silicon photomultiplier}
\newacronym{spad}{SPAD}{single photon avalanche diode}
\newacronym{pd}{PD}{photodetector}
\newacronym{rx}{Rx}{receiver}
\newacronym{tx}{Tx}{transmitter}
\newacronym{tia}{TIA}{trans-impedance amplifier}

\newacronym{pdf}{PDF}{probability density function}
\newacronym{e2e}{E2E}{end-to-end}
\newacronym{jonswap}{JONSWAP}{joint north sea wave project}
\newacronym{pm}{PM}{Pierson–Moskowitz}
\newacronym{iout}{IoUT}{internet of underwater things}
\newacronym{leo}{LEO}{low-earth orbit}
\newacronym{hn}{HN}{Hall-Novarini}
\newacronym{spf}{SPF}{scattering phase function}
\newacronym{hg}{HG}{Henyey-Greenstein}
\newacronym{ff}{FF}{Fournier-Forand}
\newacronym{psd}{PSD}{power spectral density}
\newacronym{aodn}{AODN}{australian open data network}
\newacronym{ittc}{ITTC}{International Towing Tank Conference}

\newacronym{nep}{NEP}{noise equivalent power}
\newacronym{snr}{SNR}{signal-to-noise ratio}
\newacronym{nrz}{NRZ}{non-return to zero}

\newacronym{rmse}{RMSE}{root mean square error}
\newacronym{mae}{MAE}{mean absolute error}
\newacronym{cdf}{CDF}{cumulative density function}	

\begin{document}

\title{End-to-End Optical Propagation Modeling for Water-to-Air Channels under Sea Surface and UAV Effects}

\author{M. Nennouche, M.A. Khalighi, A.A. Dowhuszko, and D. Merad
\thanks{This work was partly supported by the French PACA (Provence, Alpes, C\^ote d'Azur) Regional Council and the \'Ecole Centrale M\'editerran\'ee, Marseille, France. It was also based upon work from COST Action CA19111 (European Network on Future Generation Optical Wireless Communication Technologies, NEWFOCUS), supported by COST (European Cooperation in Science and Technology).}
\thanks{M. Nennouche is with Aix-Marseille University, CNRS, Centrale Med, LIS, Marseille, France (e-mail: mohamed.nennouche@centrale-med.fr).}
\thanks{M.A. Khalighi is with Aix-Marseille University, CNRS, Centrale Med, Fresnel Institute, Marseille, France (e-mail: ali.khalighi@fresnel.fr)}
\thanks{A.A. Dowhuszko is with Department of Information and Communication Engineering, Aalto University, Espoo, Finland (e-mail: alexis.dowhuszko@aalto.fi)}
\thanks{D. Merad is with Aix-Marseille University, CNRS, LIS, Marseille, France (e-mail: djamal.merad@lis-lab.fr)}
\thanks{Corresponding author : M.A Khalighi}}



\maketitle

\begin{abstract}
Underwater observatories have recently emerged as an efficient solution for marine biodiversity monitoring. The primary objective of this work is to enable efficient and cost-effective data muling from underwater sensors by investigating the use of optical wireless communications to transmit data from the underwater sensors to an aerial node close to the water surface, such as an unmanned aerial vehicle (UAV). More specifically, we utilize a direct water-to-air (W2A) optical communication link between the sensor node equipped with an LED emitter and the UAV equipped with an ultra-sensitive receiver, i.e., a silicon photo-multiplier. As a main contribution, we develop a comprehensive Monte Carlo–based ray-tracing algorithm to characterize this complex channel. This framework rigorously incorporates the impact of air bubbles modeled through the Mie scattering theory, a realistic sea surface representation derived from the JONSWAP spectrum, and an analytical derivation of the channel loss resulting from UAV instability under wind-induced perturbations. Furthermore, we conduct a comprehensive analysis of the W2A channel, examining the influence of key parameters such as wind speed, transmitter configurations, and receiver characteristics. The end-to-end performance evaluation demonstrates the practical feasibility of the proposed approach, achieving a bit-error rate of $10^{-3}$ at a data rate of $1$~Mbps for a transmitter depth of ~$47$\,m and wind speeds up to $13$~m/s.
\end{abstract}


\section{Introduction}
\IEEEPARstart{T}{here} has been a growing interest in the underwater world over the past decades, driven by applications such as environmental monitoring, infrastructure inspection, and resource exploration and exploitation. This interest has given rise to the concept of the \gls{iout} \cite{Jahanbakht-CST-2021}, which focuses on gathering data from underwater sensor nodes and transmitting it to remote locations for follow-up processing and storage. Achieving high-speed, low-latency, and reliable communication in such challenging environments has recently become a focal point of intensive research \cite{Khalighi-ICTON-2014, Kaushal-Access-2016, Sun-JLWT-2020}.

Given the fundamental differences in signal propagation through air and water media, different wireless technologies can be used, e.g., \gls{rf} communications in air and acoustic/optical wireless communications in water. 
To bridge these two propagation media, i.e., the underwater and the air channels, most of the previously reported approaches in the literature have considered the use of a relay node positioned on the water surface, such as a buoy or an \gls{asv} \cite{Ijeh-JOCN-2022}. This way, data from underwater sensors is transmitted to the surface node, either directly, or via an underwater drone or \gls{auv} when the sensors are located at relatively large depths \cite{Mahmoodi-TCOM-2022}. The surface node then relays the data through an \gls{rf} link to an \gls{uav} or a \gls{leo} satellite, for instance \cite{Gupta-Access-2019}.

Recently, a new approach has been considered in the literature, which consists of direct communication through the sea surface, known as \gls{w2a} communication \cite{Luo-Sensors-2022, Chen-JLT-2022}. This method represents a viable alternative to relay-based approaches in numerous relevant application scenarios. 
For instance, in the context of coral reef monitoring, this approach is particularly well-suited since coral reefs are typically located in shallow waters (at depths smaller than $50~\mathrm{m}$)\footnote{Some examples include the New Caledonian barrier reef, the Loyalty Islands lagoons, the Mayotte coral reefs, the R\'eunion lagoons, etc.}, due to their dependence on microscopic algae (Zooxanthellae) with which they live in symbiosis not far from the coast, at typically less than $30 ~\mathrm{km}$ distance \cite{Bouchet-BJLS-2002, Morgan-Nature-2016}. 
By utilizing \gls{w2a} communication, data collection can be achieved without the need for expensive underwater drones or \glspl{asv}, or the deployment of surface buoys, which often require anchoring that could disrupt and damage the delicate ecosystem under study.
\subsection{Underwater Wireless Optical Communications}
\begin{figure}[!t]
    \centerline{\includegraphics[width=0.45\textwidth]{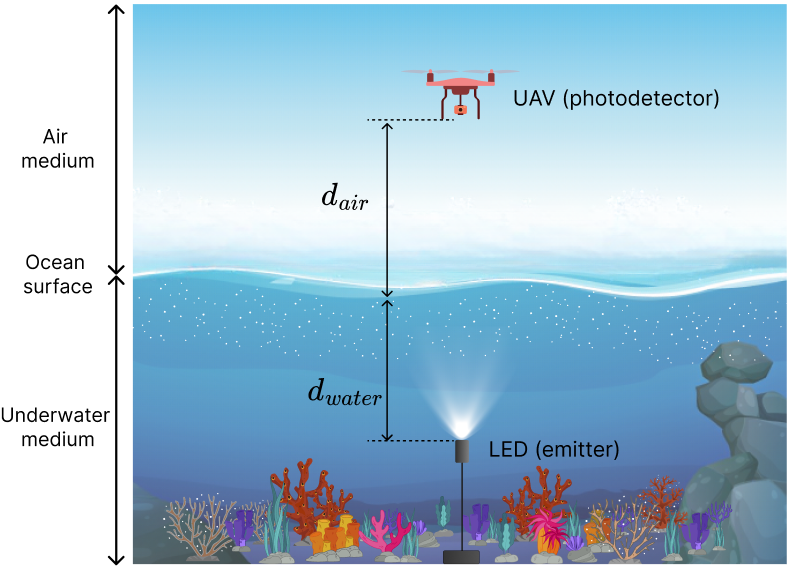}}
    \caption{\small Illustration of the considered W2A wireless optical link }
   \label{fig:model_illustration}
\end{figure}
Underwater communications have traditionally been dominated by acoustic links, primarily due to the low attenuation of acoustic signals in water, enabling communication over large ranges of up to several kilometers \cite{Ali-ACME-2020, Jiang-TWC-2023}. 
However, the limited communication bandwidth and slow propagation speed of acoustic signals result in very low data rates and significant latency. In addition, several studies have demonstrated the harmful effect of acoustic waves on underwater ecosystems \cite{Popper-JFB-2019}. As a result, there has been increasing interest in \gls{uwoc}, which offer substantially higher data rates and lower latency \cite{Kaushal-Access-2016, Khalighi-CRC-2017}, as well as a lower environmental impact \cite{Bassis-FF-2021}, thus making them more suitable for various \gls{iout} applications, particularly in environmental monitoring. These advantages, however, come at the cost of reduced link range, typically limited to a few tens of meters in clear waters \cite{Sun-JLWT-2020}. This range limitation is attributed to several impairments, including absorption, scattering, pointing errors, background noise, and oceanic turbulence \cite{Hamza-OpEX-2016, Elamassie-TCOM-2020, Ijeh-JOCN-2022}.

Here, to further reduce the potential impact on the marine ecosystem, it is crucial to limit the intensity of emitted optical signals, which can be based on using \glspl{led} or \glspl{ld}. Indeed, excessively high transmit optical power could negatively impact the surrounding flora and fauna \cite{Chen-JLT-2022}. Then, to maintain reliable communication over the typical link ranges from tens to hundred meters, it is imperative to use highly sensitive \glspl{pd} at the \gls{rx} side. In this study, we explore the use of \glspl{sipm}, also known as \glspl{mppc}, which are arrays of \glspl{spad}, that offer key advantages including very high internal gain, ease of implementation, and mechanical robustness \cite{Khalighi-PhJ-2017, Khalighi-JOE-2019, SensL-Intro-SiPM-2011}.

\subsection{Communication Across the Sea Surface}
The direct \gls{w2a} optical wireless communication approach introduces additional complexities to achieving reliable link establishment, primarily due to the dynamic and random nature of the sea surface. 
These time-dependent variations of the sea surface are influenced by numerous uncontrollable environmental factors, such as surface wind speed and underwater topography. These factors affect the optical link by causing intensity losses as well as reflection and refraction of the optical beam as it traverses the \gls{w2a} interface.
Furthermore, near the surface, wave motion generates a population of bubbles of varying sizes. These bubbles contribute to additional refraction and reflection effects and induce oceanic turbulence, which may eventually compromise the reliability of the optical wireless link.

A critical aspect in this context is the accurate modeling of the \gls{w2a} wireless channel, which is essential for designing and evaluating the performance of the communication link between the underwater sensor node and the flying \gls{uav}. While there is a rich literature on channel modeling for \gls{uwoc} and free-space optical communication \cite{Cochenour-JOE-2013, Gabriel-JOCN-2012, Elamassie-TCOM-2020, Khalighi-COMST-2014}, realistic modeling of the effect of the sea surface variations and the impact of air bubbles, surface waves, and \gls{uav} instability on the intensity of the received signal requires particular attention. Note, realistic sea surface modeling has been extensively studied by the oceanography community with a variety of models proposed so far. 
\subsection{Use of UAVs in Open Ocean Environments}

In recent years, \glspl{uav} have emerged as versatile and cost-effective platforms for advancing maritime wireless communication systems \cite{Nomikos-OJCOMS-2022}. Their rapid deployment capability and mobility make them particularly suitable for open-ocean environments, where traditional infrastructures are either unavailable or prohibitively expensive. \glspl{uav} can establish links with \glspl{asv}, floating relay buoys \cite{Pottoo-SII-2022, Nomikos-OJCOMS-2022}, or directly with underwater objects, thereby supporting critical applications such as deep-sea exploration, environmental monitoring, and disaster-response operations.

Nevertheless, the integration of \glspl{uav} into maritime communication networks introduces significant challenges. Their performance is highly sensitive to meteorological conditions, especially wind, which induces vibrations, link misalignment, and fluctuations in the angle of signal reception \cite{Pottoo-SII-2022, Liu-TVT-2022}. Although these effects have been investigated in the literature for both \gls{rf} and optical wireless systems \cite{Agheli-JLT-2021, Pottoo-SII-2022, Liu-TVT-2022, Nie-SECON-2022}, a rigorous characterization of \gls{uav} dynamics under different levels of wind disturbances, as well as their impact on optical W2A links remains largely unexplored. To the best of our knowledge, this study is the first to propose a comprehensive modeling framework capturing \gls{uav} vibration behavior as a function of wind speed in the context of optical \gls{w2a} communication.

\subsection{Motivations and Contributions}
This work focuses on the context of coral reef monitoring, where environmental sensing data collected by an underwater sensor are transmitted via an \gls{w2a} optical link to a flying \gls{uav}.\footnote{When a wireless sensor network is deployed, one node typically serves as the coordinator, collecting data from all sensor nodes and transmitting them to the aerial node. The \gls{w2a} link under consideration will then correspond to that between the coordinator node and the \gls{uav}.} This application scenario is depicted in Fig.\,\ref{fig:model_illustration}. 
To simplify the design of the underwater node, the \gls{tx} at depth $d_{\mathrm{water}}$ employs an \gls{led} optical source, while the \gls{uav}, positioned at height $d_{\mathrm{air}}$ above the sea surface, is equipped with an \gls{sipm}-based optical \gls{rx}. As a very sensitive photo-detector is used, data muling from the sensor to the \gls{uav} is assumed to be performed at night to minimize the effect of background noise \cite{Hamza-WACOWC-2019}.

The practical implementation of such intermediate systems is severely hampered by the extremely dynamic nature of the propagation environment. Motivated by the lack of comprehensive models in the literature that jointly account for these severe environmental impairments, we conduct an in-depth analysis of the \gls{w2a} channel. This characterization accounts for the absorption and scattering effects caused by underwater particles and air bubbles, as well as the influence of surface waves and \gls{uav} instability under varying wind speeds.

To account for the effect of surface waves, we adopt the \gls{jonswap} model \cite{JONSWAP-1973}, which is widely accepted within the oceanography research community \cite{Ryabkova-JGRO-2019, Lee-JMSE-2022, Rueda-JWPCOE-2020}. In particular, this model incorporates a key parameter, i.e., the distance between the area of study (with wind activity at the sea surface) and the seashore. This parameter is particularly relevant for coral reef studies, as these natural environments are typically located a few kilometers from the coasts \cite{Bouchet-BJLS-2002}. On the other hand, in order to account for the effect of air bubbles, we propose to use the \gls{hn} model \cite{Keiffer-JASA-1995, Ainslie-JASA-2005}, which describes the distribution of the bubble population as a function of depth. Additionally, we apply the Mie theory \cite{Mobley-CRC-1994} to quantify the scattering effects introduced by these air bubbles. Finally, as an important consideration, we incorporate a study of the effect of \gls{uav} instability induced by turbulent wind conditions, represented through the Dryden model \cite{Kachroo-TAP-2021, Dryden-1990}. These considerations allow a comprehensive and realistic characterization of the \gls{w2a} channel, which is critical for link design and performance evaluation.

To characterize the \gls{w2a} channel, we employ Monte Carlo simulations \cite{Gabriel-JOCN-2012} to model photon propagation across the wireless link, encompassing both underwater and aerial segments. Subsequently, we incorporate the channel loss induced by \gls{uav} instability, for which we derive a closed-form expression. To the best of our knowledge, this study is the first to provide a closed-form formulation of \gls{uav} tilt angle variations and the corresponding loss in the context of optical wireless communications.

Ultimately, we examine the impact of various system parameters, including the \gls{led} beam divergence and the \gls{fov} of the \gls{rx}, and the \gls{e2e} link performance. This performance is evaluated in terms of the \gls{ber}, providing insights into the practical feasibility of the proposed approach.

The key contributions of this work can be summarized as follows:
\begin{itemize}
    \item Development of a comprehensive Monte Carlo ray-tracing framework to accurately model the \gls{w2a} optical channel, specifically tailored for realistic shallow-water coral reef monitoring scenarios;
    \item Integration of advanced physical models to quantify complex environmental impairments. This includes employing the \gls{jonswap} spectrum to capture near-shore sea surface dynamics, and combining the \gls{hn} distribution with Mie theory to rigorously evaluate the scattering effects of underwater air bubbles;
    \item {Derivation of a closed-form analytical expression to characterize the channel losses induced by \gls{uav} orientation fluctuations. This includes a mathematically tractable approximation of the \gls{uav} inclination angle under turbulent wind conditions, modeled via the Dryden wind spectrum;
    \item Extensive end-to-end link performance analysis}, evaluating the impact of wind speed, operational \gls{tx} depth, and data rates on the \gls{ber}. The results conclusively demonstrate the feasibility and reliability of \gls{w2a} optical communications for marine data muling applications.
\end{itemize}

The remainder of this paper is organized as follows:  Section\,\ref{sec:Related_works} presents a brief state-of-the-art on \gls{w2a} optical channel modeling and link design, whereas, Section\,\ref{sec:System_Model} describes in detail the proposed channel modeling to account for several factors such as underwater beam absorption and scattering, random sea surface, air bubbles, and \gls{uav} instability. Afterwards, Section\,\ref{sec:uav_instability} presents our new approach to modeling losses resulting from drone instability. Subsequently, Section\,\ref{sec:Monte_Carlo} details the Monte Carlo algorithm that has been developed to model photon propagation, while Section\,\ref{sec:Results} presents the numerical results, focusing on channel modeling and performance evaluation of the wireless link. Finally, Section\,\ref{sec:concluson} draws the main conclusions of this work.

\section{State-of-the-Art on W2A Optical Communications}\label{sec:Related_works} 
In recent years, there has been growing interest in \gls{w2a} wireless optical communication, focusing on both channel modeling and link performance evaluation. Several recent studies have also conducted experiments to assess \gls{w2a} link performance, utilizing indoor test-beds such as water tanks or pools, as well as outdoor setups in harbor environments.

For instance, in \cite{Sun-Express-2019}, Sun \textit{et al.} experimentally demonstrated a high-speed \gls{w2a} link using an ultraviolet \gls{led} and different modulation schemes, and analyzed the link performance in terms of transmission rate under the conditions of both perfect beam alignment and slight misalignment (on the order of a few centimeters) in the presence of waves of up to $15$~mm in height. 
More recently, Lin \textit{et al.} proposed in \cite{Lin-Photonics-2024} a prototype based on \gls{mimo} technique, using $4$ blue \glspl{led} and $4$ \glspl{apd}, as well as spatial optical filtering to reduce the impact of background radiations. Through a series of experiments conducted in a water tank and in a swimming pool, they demonstrated the performance improvement achieved using the \gls{mimo} technique and error correction coding.

Nevertheless, in the environmental monitoring application under consideration, conducting experiments that accurately reflect the real-world conditions of the open sea, with waves reaching several meters in height, is highly challenging. Therefore, simulation-based approaches, such as those based on Monte Carlo algorithms, represent the most suitable way for channel characterization and modeling. A critical aspect of these simulations is the realistic modeling of the air-water interface.
One established model is the \gls{pm} model \cite{Pierson-JGR-1964}, which, for instance, was used in \cite{Lin-ICCC-2020} to study the spatial distribution of the \gls{w2a} channel loss and its temporal evolution under varying wind speeds. Using this model, system performance with multiple \glspl{tx} and/or \glspl{rx} was analyzed and experimentally validated in \cite{Lin-JLT-2022}. Another notable approach for the air-water interface modeling is the three-dimensional ECKV\footnote{Standing for the initials of the authors' names: Elfouhaily, Chapron, Katsaros, and Vandemark.} model \cite{Elfouhaily-JGRO-1997}, which was used in \cite{Angara-Access-2024} to investigate the \gls{w2a} channel. The authors of \cite{Angara-Access-2024} also studied the impact of air bubbles in the underwater channel and solar background noise considering a laser-based \gls{tx} and an \gls{apd}-based \gls{rx}.

Both \gls{pm} and ECKV models offer realistic sea-surface modeling but are primarily applicable to systems deployed in the open sea, i.e., at typically several hundred kilometers from the sea shore. To the best of our knowledge, to date, no prior studies have investigated the optical \gls{w2a} channel in near-shore environments, where the \gls{jonswap} model becomes particularly relevant for capturing the unique dynamics of such scenarios.
\section{W2A Channel Modeling}\label{sec:System_Model}
In this section, we develop appropriate mathematical models for the four key components that need to be considered for modelling the \gls{w2a} channel, i.e., the underwater medium, the \gls{w2a} interface, the effect of air bubbles near the sea surface, and the instability of the \gls{uav} due to the wind speed. Accordingly, we present the mathematical foundations underlying the calculation of the associated losses.

The \gls{w2a} channel loss $h$ can be modeled as:
\begin{equation}
    h = h_\alpha\, h_\mathrm{MC},
\end{equation}
where, $h_\alpha$ and $h_\mathrm{MC}$ denote the orientation-induced loss resulting from \gls{uav} tilting under wind effects, and the channel loss accounting for the ensemble of attenuation in water, air bubbles effect, and sea surface fluctuations, respectively. In Section~\ref{sec:uav_instability}, we derive the analytical expression of $h_\alpha$, and subsequently, in Section~\ref{sec:Monte_Carlo}, we introduce our Monte-Carlo simulator used for estimating $h_\mathrm{MC}$.
\subsection{Modeling Optical Propagation in Water}\label{Subsec:Model-water}
Before reaching the sea surface, the emitted photons from the \gls{tx} pass through the underwater medium, introducing absorption and scattering \cite{Mobley-CRC-1994}. 
As we consider the sensor node for coral reefs monitoring to be at a relatively small depth (typically lower than $50$\,m), we logically assume that the distribution of underwater particles is almost uniform \cite{Lesser-Springer-2007}. Additionally, when working at relatively shallow waters, we reasonably consider the effects of oceanic turbulence caused by marine currents or temperature and salinity gradients to be negligible (a detailed justification of this assumption is provided in Appendix\,\ref{app:temp_sal}).

As will be detailed in Section\,\ref{sec:Monte_Carlo}, we use Monte Carlo simulations to model photon transport in the underwater channel. 
Photon scattering modeling will rely on the \gls{spf} $\widetilde{\beta}(\theta)$, which provides the scattering angle $\theta_s$ of a photon after interaction with a particle. To obtain samples of $\theta_s$, a random variable $\xi_\beta$, uniformly distributed between $0$ and $1$, denoted here by ${\cal{U}}(0,1)$, will be generated to solve the following equation:
\begin{equation}
    \label{theta_s_beta}
    \xi_\beta = 2\pi\int_0^{\theta_s} \widetilde{\beta}(\theta)\sin\theta \, d\theta.
\end{equation}
The most commonly used \gls{spf} is the \gls{hg} function \cite{Gabriel-JOCN-2012,Lin-JLT-2022, Sahoo-Optics-2022, Qin-Optics-2022} described by:
\begin{equation}
    \label{HG_PDF}
    \widetilde{\beta}_{\mathrm{HG}}(\theta_s, g_\mathrm{HG}) = \frac{1 - g_\mathrm{HG}^2}{4\pi\left(1 + g_\mathrm{HG}^2 - 2g_\mathrm{HG}\cos\theta_s\right)^{3/2}},
\end{equation}
where $g_\mathrm{HG}$ is the mean cosine angle in all directions.
The interest of the \gls{hg} function is mainly due to the simplicity of its formulation and the ease of finding the scattering angle $\theta_s$ using the inverse transform sampling method. That is, using a random variable $\xi_\mu$, of distribution ${\cal{U}}(0,1)$, we obtain \cite{Gabriel-JOCN-2012},
\begin{equation}
    \label{HG_cos_theta-1}
    \mu_{\mathrm{HG}} = \frac{1}{2g_\mathrm{HG}}\left[1 + g_\mathrm{HG}^2 - \left(\frac{1 - g_\mathrm{HG}^2}{1 - g_\mathrm{HG} + 2\,g_\mathrm{HG}\,\xi_\mu}\right)^2\right] .
\end{equation}
Then, the samples of the scattering angle $\theta_s$ are obtained as
\begin{equation}
    \label{HG_cos_theta-2}
 \theta_s = \arccos\mu_{\mathrm{HG}}.
\end{equation}
However, the \gls{hg} function lacks precision for small ($\lesssim20^\circ$) and large scattering angles ($\gtrsim120^\circ$), when compared to the experimental data \cite{Mobley-AOptics-2002, Gabriel-JOCN-2012, Bohren-book88}. Several alternative \gls{spf} models have been proposed in the literature, where one of the most commonly used is the \gls{ff} function \cite{Liu-OptEng-2017, Fournier-SPIE-1999, Haltrin-IEEE-1998}, given by (\ref{FF_scat_phase_function}), that we will consider in this paper.
\begin{multline}
    \label{FF_scat_phase_function}
    \widetilde{\beta}_{\mathrm{FF}}(\theta_s, \mu_{\mathrm{FF}}, n_s) = \frac{1}{4\pi(1-\delta_{\theta_s})^2\delta_{\theta_s}^v}\bigg\{v(1-\delta_{\theta_s})-(1-\delta_{\theta_s}^v) \\  +\big[\delta_{\theta_s}(1-\delta_{\theta_s}^v)-v(1-\delta_{\theta_s})\big]\sin^{-2}(\frac{\theta_s}{2})\bigg\} \\ + \frac{1 - \delta_\pi^v}{16\pi(\delta_\pi - 1)\delta_\pi^v}(3\cos^2\theta_s - 1).
\end{multline}

Here, $n_s$ and $\mu_{\mathrm{FF}}$ represent the refractive index of the particle and the inverse power of Junge distribution\footnote{The Junge distribution is a special case of a power law describing the size distribution of underwater particles \cite{OceanOptics_Website}.}, respectively \cite{Liu-OptEng-2017}, and

\begin{equation}
    v = \frac{3 - \mu_{\mathrm{FF}}}{2} ~ , ~\delta_{\theta_s} = \frac{4}{3(n_s-1)^2}\sin^2\left(\frac{\theta_s}{2}\right).
\end{equation}

Obviously, $\widetilde{\beta}_{\mathrm{FF}}$ is a more complex function than $\widetilde{\beta}_{\mathrm{HG}}$. Furthermore, since the angle $\theta_s$ cannot be directly expressed as a function of $(\mu_{\mathrm{FF}}, n_s)$, we have to calculate the integral in (\ref{theta_s_beta}) numerically. Here, to reduce the computational complexity, we use an approximation of this integral as follows \cite{Liu-OptEng-2017}.

\begin{multline}
    \label{theta_s_FF}
    \xi_{\mathrm{FF}} = 2\pi\int_0^{\theta_s} \widetilde{\beta}_{\mathrm{FF}}(\theta)\sin\theta d\theta \\ 
    \approx  \frac{1}{(1-\delta)\delta^v}\left[(1 - \delta^{v+1}) - (1 - \delta^v)\sin^2\left(\frac{\theta_s}{2}\right)\right] \\
    + \frac{1}{16\pi}\frac{1 - \delta_\pi^v}{(\delta_\pi - 1)\delta_\pi^v}\cos\theta_s\sin^2\theta_s,
\end{multline}
where $\xi_{\mathrm{FF}}$ is a random variable of distribution ${\cal{U}}(0,1)$. The scattering angle $\theta_s$ is then obtained from $\xi_{\mathrm{FF}}$ through numerical approximation.

\begin{figure}
    \centering
    \includegraphics[width=0.485\textwidth]{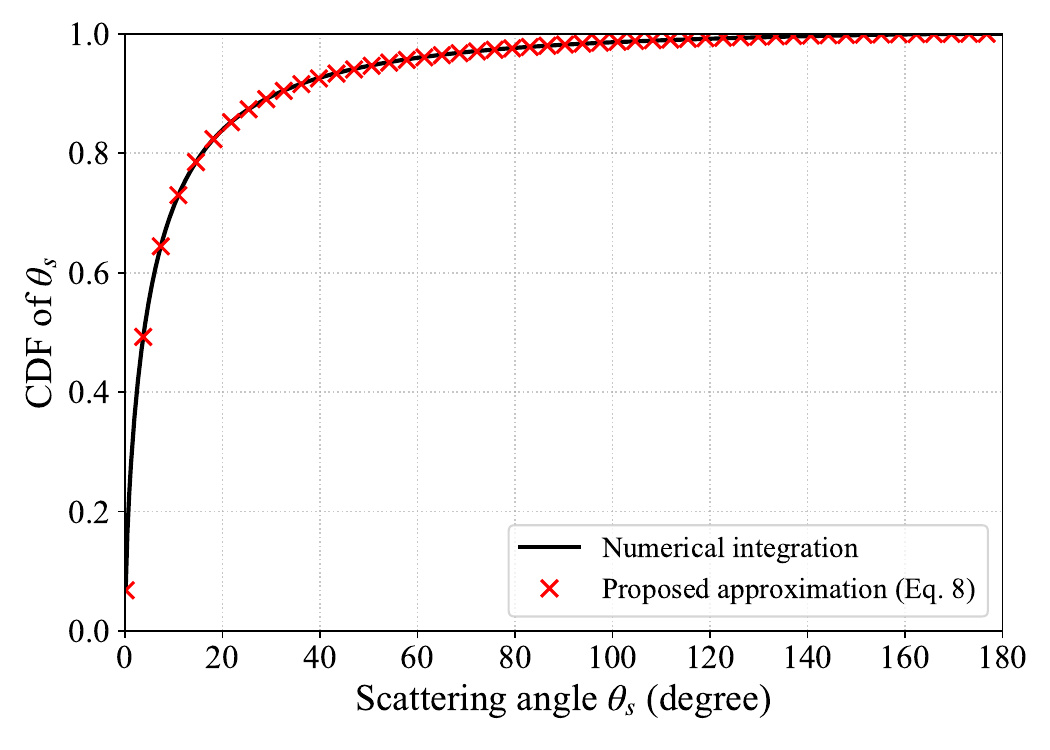}
    \caption{Comparison between the exact numerical integration and the proposed approximation in \eqref{theta_s_FF}.}
    \label{fig:cdf_ff_approximation}
\end{figure}

To evaluate the accuracy of this approximation and its potential impact on the photon propagation model, we compare the \gls{cdf} of $\theta_s$ obtained via the exact numerical integration of \eqref{theta_s_beta} using $\widetilde{\beta}_{\mathrm{FF}}(\theta_s)$ with that of the proposed approximation in \eqref{theta_s_FF}. For this comparison, we adopt parameters representative of clear ocean waters \cite{Mobley-AOptics-2002}, specifically setting the refractive index to $n_s = 1.1$ and the slope parameter to $\mu_{\mathrm{FF}} = 3.5835$. As shown in Fig.~\ref{fig:cdf_ff_approximation}, the approximate \gls{cdf} closely matches the exact numerical evaluation across the entire angular range ($0^{\circ}$ to $180^{\circ}$). To quantitatively assess the goodness of fit, we evaluate the \gls{rmse}, the \gls{mae}, as well as the Kolmogorov-Smirnov statistic, denoted as $D_{\mathrm{KS}}$, which represents the maximum absolute difference between the two cumulative distributions \cite{Wasserman-book-2004}. Based on our analysis, we obtain an \gls{rmse} of $6.28 \times 10^{-4}$, a \gls{mae} of $5.33 \times 10^{-4}$, and a $D_{\mathrm{KS}}$ of $9.69 \times 10^{-4}$. These remarkably tight error margins demonstrate that the proposed approximation significantly reduces the computational overhead inherent to our Monte Carlo-based ray-tracing simulator without compromising the physical accuracy of the end-to-end channel model.
\subsection{Modeling the W2A Interface}\label{Subsec-W2A-Interface}
Mathematical modeling of sea wave shapes remains a challenging open problem due to the complexity and interplay of numerous contributing factors, such as underwater currents, atmospheric pressure, proximity to shore, surface winds, etc. Among these, surface wind speed has a particularly significant effect, as it plays a crucial role in determining sea surface elevation. So far, the model developed by Cox and Munk in \cite{Cox-JOSA-1954} has been widely used in the literature (e.g. in \cite{Sahoo-Optics-2022, Qin-Optics-2022}), primarily due to the simplicity of its formulation. 
This model proposes the sea surface slope in the form of a probabilistic distribution whose variance is dependent on the wind speed. Unfortunately, this purely random approach does not take into account the spatial and temporal correlation that exists in practice between adjacent points of the sea surface. Consequently, other models based on the frequency and angular spectrum of waves have been proposed \cite{Cavaleri-ProgressOceano-2007} such as the \gls{pm} \cite{Pierson-JGR-1964} and the \gls{jonswap} models \cite{JONSWAP-1973}, previously introduced in Section~\ref{sec:Related_works}.

The \gls{jonswap} model is widely accepted in oceanography for typical wind speeds (excluding extreme cases such as hurricanes) and is considered as an evolution of the \gls{pm} model.
It incorporates a so-called \textit{fetch} parameter $F$, which represents the distance over which the wind interacts with the water surface and can be interpreted as the distance from the seashore. In fact, this parameter adjusts wave heights by modifying the amplitudes of the frequency spectrum. The incorporation of this parameter is particularly relevant in the context of this work, which aims to provide an effective solution for monitoring coral reef ecosystems.

Based on the \gls{jonswap} model, we represent the sea surface elevation $z(x,y,t)$ as a sum of sinusoids, as given by (\ref{water_surface}), where $(x, y)$ and $t$  denote the spatial coordinates and time, respectively \cite{Dong-Sensors-2020}.
\begin{equation}
    \label{water_surface}
    z(x,y,t) = \sum_{i=1}^{M}\sum_{j=1}^{N} a_{ij}\cos(\omega_it-k_ix\cos\theta_j-k_iy\sin\theta_j+\epsilon_{ij})
\end{equation}
Here, $M$ and $N$ represent the numbers of frequencies and directional angles considered in the wave spectrum, respectively. Also, $a_{ij}$, $\omega_i$, $k_{i}$, and $\theta_j$ stand for the amplitude of the wave components, wave angular frequency, wave number, and directional angle, respectively. Finally, $\epsilon_{ij}$ denotes the initial phase of each wave component, which follows the distribution ${\cal{U}}(0,2\pi)$.
In ocean wave studies, the values of $\omega_i$ are calculated as a function of the wave number $k_i$ as \cite{OceanOptics_Website}:
\begin{equation}
\omega_i^2 = g\, k_i\, \tanh(d_0k_i),
\end{equation}
where $d_0$ represents the sea depth. 
For $d_0 \rightarrow \infty$, we use the approximation $\tanh(d_0k_i) \rightarrow 1$, which is commonly accepted 
when the sea depth significantly exceeds the wavelength of the sea waves under study, as is the case in this work. As a result,
\begin{equation}
\omega_i^2 \approx g\, k_i.
\end{equation}
On the other hand, the amplitude of each component of the wave $a_{ij}$ and the \gls{jonswap} wave spectrum $S(\omega)$ are related through the following equation,
\begin{equation}
    \label{aij}
    a_{ij} = \sqrt{2\, S(\omega)D(\theta)\, \Delta\omega\, \Delta\theta},
\end{equation}
where $D(\theta)$ represents the directional spreading function, which governs the direction of wind evolution \cite{Dong-Sensors-2020}, and $\Delta\omega$ and $\Delta\theta$ denote the sampling steps in the frequency and angular domains, respectively. 
Several models have been proposed for $D(\theta)$ in the literature \cite{Lin-CRC-2008}. In this work, we consider the widely-used model proposed by the \gls{ittc} association \cite{Dong-Sensors-2020,Song-ISOPE-2018, Lee-OE-2006}, which is described by:
\begin{equation}
    \label{spreading_function}
    D(\theta) = \frac{2}{\pi}\cos^2\theta, \quad |\theta|\leq \frac{\pi}{2}.
\end{equation}
Furthermore, the wave spectrum in (\ref{aij}) is calculated as \cite{JONSWAP-1973}:
\begin{equation}
    \label{JONSWAP_spectrum}
    S(\omega) = \frac{\alpha \, g^2}{\omega^5}\exp\left[-\frac{5}{4}(\frac{\omega_0}{\omega})^4\right]\ \gamma^r,
\end{equation}
where,
\begin{equation}
\gamma = 3.3~,~\alpha = 0.075\, X^{-0.22},
\end{equation}
and $X = g\, F/U_{10}^2$, with $F$ being the fetch parameter which is set in this work to $30$~km, and $U_{10}$ the average wind speed at $10$\,m above the sea surface.
Also, 
\begin{equation}
    r = \exp\left[- \frac{(\omega-\omega_0)^2}{2\sigma^2\omega_0^2}\right], 
\end{equation}
where
\begin{equation}
    \label{sigma_JONSWAP}
    \sigma= 
    \begin{cases}
        0.07 & , \quad\omega \leq \omega_0, \\ 
        0.09 & , \quad\omega>\omega_0,
    \end{cases} 
\end{equation}
and $\omega_0$ represents the peak angular speed of the spectrum, calculated as follows.
\begin{equation}
    \label{omega_0_JONSWAP}
    \omega_0 = 22\, \frac{g}{U_{10}}\, X^{-0.33}
\end{equation}
\subsection{Modeling the Effect of Air Bubbles}\label{Subsec-effect-air-bubbles}
As mentioned earlier, the movement of the sea surface with waves generates a population of bubbles near the surface \cite{Keiffer-JASA-1995, Angara-Access-2024} that should be taken into account in channel modeling as their impact can be significant. As explained, in this work, we use the widely accepted \gls{hn} model \cite{Keiffer-JASA-1995} to account for the effect of air bubbles on optical wave propagation \cite{Angara-Access-2024, Sahoo-Optics-2022}. For the sake of simplicity, we only take into account the scattering effect of bubbles on the optical beam, as the propagation loss within the air bubble is rather negligible. In other words, the effect of air bubbles is evaluated considering the bubble scattering coefficient by:
\begin{equation}
    \label{HN_b_bub}
    b_{\mathrm{bub}}(z) = N_b(z)\, Q_{sca} \, \Psi,
\end{equation}
where $N_b(z)$, $Q_{sca}$, and $\Psi$ denote the density of bubbles as a function of depth $z$, the mean scattering efficiency (here equal to $2.0$, considering particles of radius larger than $1\,\mu$m \cite{Zhang-AOptics-1998}), and the mean geometric cross-sectional areas of the bubble populations, respectively.
To calculate $N_b(z)$, we integrate the bubble size distribution $n(r,z)$ over $r$, which denotes the radius of the bubbles. For the \gls{hn} model, this latter is given by:
\begin{equation}
    \label{HN_n_z}
    n(r,z) = 1.6\times 10^{10}\ G(r,z)\, \left(\frac{U_{10}}{13}\right)^3\ \exp\left(-\frac{z}{L(U_{10})}\right).
\end{equation}
Here, $L(U_{10})$ is the so-called e-folding distance in units of meter, given by:
\begin{equation}
    \label{HN_L_U}
    L(U_{10})= \begin{cases}0.4 &, \quad U_{10} \leq 7.5\ \mathrm{m}/\mathrm{s}, \\ 0.4+0.115\left(U_{10}-7.5\right) &, \quad U_{10}>7.5\ \mathrm{m}/\mathrm{s}.\end{cases}
\end{equation}
Also, $G(r, z)$ in (\ref{HN_n_z}) is the factor that governs the dependence of the spectrum on radius $r$ at a given depth $z$ and is given by \cite{Keiffer-JASA-1995}:
\begin{equation}
    \label{HN_G_z}
    G(r, z)= \begin{cases}{\left[r_{\mathrm{ref}}(z) / r\right]^4} &,\quad r_{\min } \leq r \leq r_{\mathrm{ref}}(z), \\ {\left[r_{\mathrm{ref}}(z) / r\right]^\kappa} &,\quad r_{\mathrm{ref}}(z)<\mathrm{r} \leq r_{\max},\end{cases}
\end{equation}
where $r_{\mathrm{ref}}$ refers to the reference radius of the bubble population in the ocean, calculated using (\ref{r_ref}), and $\kappa = 4.37+(z/2.55)^2$. 
\begin{equation}
\label{r_ref}
    r_{\mathrm{ref}} = 54.4\, \mu \mathrm{m} + 1.984 \times 10^{-6}z
\end{equation}
Also, $r_{\min}$ and $r_{\max}$ in (\ref{HN_G_z}) denote the minimum and maximum bubble radii, set here to $10~\mu$m and $1$\,mm, respectively, which correspond to the limits considered in the \gls{hn} model \cite{Ainslie-JASA-2005}.
Lastly, $N_b(z)$ and $\Psi$ in (\ref{HN_b_bub}) are calculated as follows \cite{Ma-Express-2015}:
\begin{multline}
    \label{HN_N_z}
    N_b(z) = \int_{r_{\min}}^{r_{\max}}n(r,z)dr \\ \approx (1.6\times10^{10})\frac{r_{\mathrm{ref}}^4}{3\,r_{\min}^3}\bigg(\frac{U_{10}}{13}\bigg)^3\, \exp\left[-\frac{z}{L(U_{10})}\right],
\end{multline}
\begin{multline}
    \label{HN_S}
    \Psi = \int_{r_{\min}}^{r_{\max}} \frac{n(r,z)}{N_b(z)}\pi r^2 dr \\ = \int_{r_{\min}}^{r_{\max}} \frac{3\,r_{\min}^3}{r_{\mathrm{ref}}^4}\, G(r,z)\,\pi r^2dr \approx 3\pi r^2_{\min}.
\end{multline}

Now, to account for the scattering effect of air bubbles, it is important to note that their size differs significantly from that of typical particles in water, leading to fundamentally different scattering behaviors. In this paper, we employ Mie theory as the appropriate model to calculate the scattering angle $\theta_s$ of a photon following its interaction with an air bubble \cite{Ma-Express-2015, Sahoo-Optics-2022}. Based on this, we evaluate the \gls{spf} of bubbles $\widetilde{\beta}_{\mathrm{bub}}$ as follows \cite{Ma-Express-2015}:
\begin{equation}
    \label{Mie_SPF}
    \widetilde{\beta}_{\mathrm{bub}}(z, \theta) = \frac{1}{b_{\mathrm{bub}}}\int_{r_{\min}}^{r_{\max}}Q_b(r,\theta)\,\pi \,r^2\, n(r,z)dr,
\end{equation}
where $Q_b(r,\theta)$ ($\mathrm{sr}^{-1}$) is the scattering efficiency per unit solid angle in the direction $\theta$ for a single bubble with radius $r$.
\subsection{Effect of Wind on UAV Orientation}

\begin{figure}
        \centering
        \includegraphics[width=0.8\linewidth]{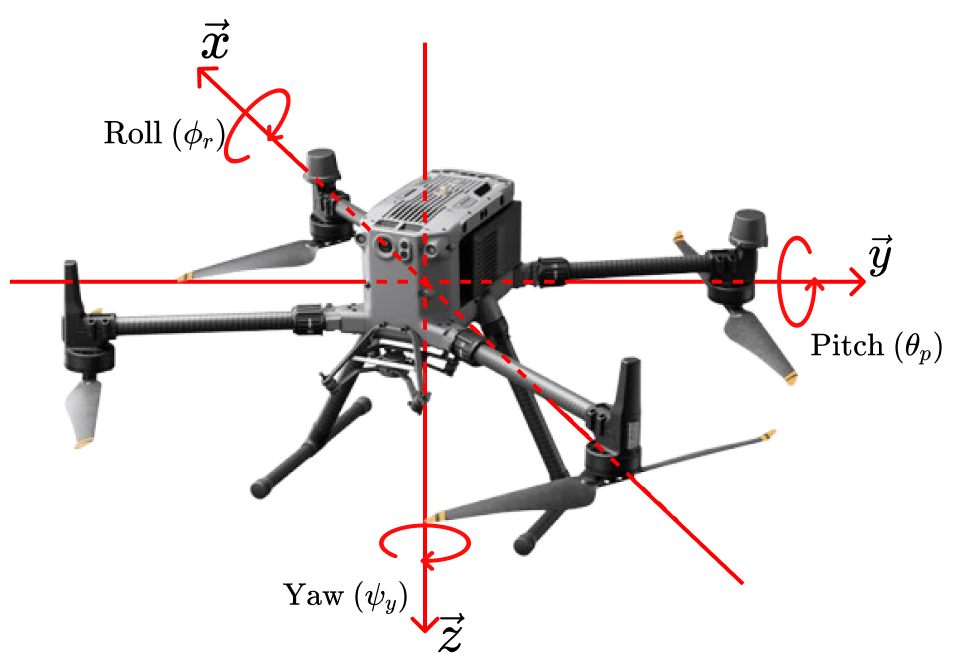}
        \caption{\small Axes and rotation angles of a typical quadcopter UAV that can be used for data muling in the considered coral reef monitoring scenario.}
        \label{fig:roll_pitch_yaw}
    \end{figure}

In maritime environments, the frequent occurrence of wind gusts and turbulent air flows can induce significant deviations in \gls{uav} orientation \cite{Wang-MeasCont-2019}. Such angular instabilities directly affect the optical beam alignment and, consequently, the reliability of the \gls{w2a} communication link. For most \glspl{uav} employed in such missions, ``flight controllers'' are used to actively compensate for wind disturbances through roll $(\phi_r)$, pitch $(\theta_p)$, and yaw $(\psi_y)$ adjustments, as illustrated in Fig.~\ref{fig:roll_pitch_yaw}.

    To model wind speed fluctuations, we adopt the Dryden wind turbulence model \cite{Dryden-1990}, which characterizes the amplitude of turbulence in each spatial direction, represented by $\widetilde{U} = (\tilde{u}_x, \tilde{u}_y, \tilde{u}_z)$. Denoting the average and instantaneous wind speeds by vectors $U = (u_x, u_y, u_z)$ and $V= (V_x, V_y, V_z)$, we have $V= U + \widetilde{U}$.
    
    To estimate the \gls{uav} deflection angle $\alpha_\mathrm{UAV}$ under wind disturbances, we adopt the approach proposed in \cite{Palomaki-JACT-2017} and validated in \cite{Wetz-AMT-2021}, leading to the following direct relationship between $\alpha_\mathrm{UAV}$ and $V$ \cite{Palomaki-JACT-2017}:
    \begin{equation}
        \label{eq:tan_alpha}
        \tan(\alpha_\mathrm{UAV}) = \frac{1}{2\,m_\mathrm{UAV}\,g}C_D\,\rho_\mathrm{air}\,A_\mathrm{UAV}V^2,
    \end{equation}
    where $\rho_\mathrm{air} = 1.293\, \text{kg/m}^3$ denotes air density, and $m_\mathrm{UAV}$, $C_D$, and $A_\mathrm{UAV}$ correspond to the mass, drag coefficient, and equivalent flat area of the \gls{uav}. Lastly, $\alpha_\mathrm{UAV}$ is related to $\theta_p$ and $\phi_r$ through the following equation \cite{Palomaki-JACT-2017}:
    \begin{equation}
        \alpha_\mathrm{UAV} = \arccos(\cos\theta_p\cos\phi_r).
    \end{equation}
    Further details on the Dryden model are provided in Appendix \ref{app:Dryden_implementation}.

\section{Statistical Modeling of UAV Instability}\label{sec:uav_instability}

Here, we introduce our methodology for modeling the impact of turbulent wind on the \gls{uav} stability based on the Dryden turbulence model \cite{Dryden-1990}. We first derive a closed-form expression for the \gls{pdf} of the \gls{uav} deflection angle, and subsequently formulate the corresponding channel loss.

\subsection{UAV Deflection Angle Modeling}
Let $\mathcal{N}(\mu_n,\sigma_n^2)$ denote a Gaussian distribution with mean $\mu_n$ and variance $\sigma_n^2$. Based on the Dryden model, we consider $V_x \sim \mathcal{N}(u_x, \sigma_x^2/\pi)$, $V_y \sim \mathcal{N}(u_y, \sigma_y^2/\pi)$, and \mbox{$V_z \sim \mathcal{N}(u_z, \sigma_z^2/\pi)$}, where $\sigma_x^2$, $\sigma_y^2$, and $\sigma_z^2$ denote the variances of wind turbulence along the three axes $x$, $y$, and $z$ (note, a detailed proof of this statement is provided in Appendix~\ref{app:distribution_proof}). 
    
To obtain the \gls{pdf} of $\alpha_\mathrm{UAV}$, we define:
\begin{equation}
    Z_j = \frac{V_j - u_j}{\sigma_j/\sqrt{\pi}} \sim \mathcal{N}(0, 1), 
    \qquad j \in \{x,y,z\},
\end{equation}
which leads to
\begin{equation}
    V^2 = \sum_{j \in \{x,y,z\}} \frac{\sigma_j^2}{\pi}(Z_j + \lambda_j)^2 
    = \sum_{j \in \{x,y,z\}} \frac{\sigma_j^2}{\pi} \chi'^2_1(\lambda_j^2),
\end{equation}
where $\lambda_j = u_j\sqrt{\pi}/\sigma_j$, and $\chi'^2_1(\lambda_j^2)$ denotes a non-central Chi-squared distribution with non-centrality parameter $\lambda_j^2$. Note, $V^2$ does not follow a standard distribution, i.e., no simple closed-form expression is available to characterize its \gls{pdf}. Consequently, we propose to approximate the \gls{pdf} of $V^2$ by a Gamma distribution, using the moment matching method with the aim of fitting the first and second order moments, i.e., $E[V^2]$ and $\mathrm{Var}[V^2]$, where $E[.]$ and $\mathrm{Var}[.]$ denote the expected value and variance, respectively. This way, the parameters $\alpha_\Gamma$ and $\beta_\Gamma$ of the corresponding Gamma distribution $\Gamma(\alpha_\Gamma, \theta_\Gamma)$ are calculated as follows:
\begin{equation}
    \alpha_\Gamma = \frac{[E[V^2]]^2}{\mathrm{Var}[V^2]}, 
    \qquad 
    \beta_\Gamma = \frac{\mathrm{Var}[V^2]}{E[V^2]},
\end{equation}
where
\begin{align}
    \begin{aligned}
        E[V^2] & = \sum_{j \in \{x,y,z\}} \frac{\sigma_j^2}{\pi}\left(1 + \lambda_j^2\right), \\
        \mathrm{Var}[V^2] & = \sum_{j \in \{x,y,z\}} \frac{\sigma_j^4}{\pi^2}\left(2 + 4\lambda_j^2\right).
    \end{aligned}
\end{align}
Consequently, the approximate \gls{pdf} of $V^2$, denoted here by $f_{V^2}(v)$, is given by:
\begin{equation}
    f_{V^2}(v) = \frac{v^{\alpha_\Gamma - 1} \exp(-v/\beta_\Gamma)}{\Gamma(\alpha_\Gamma)\,\beta_\Gamma^{\alpha_\Gamma}}.
\end{equation}

Finally, by applying the variable change of \mbox{$V^2 = \tan(\alpha_\mathrm{UAV})/K$}, with $K = \frac{C_D\, \rho_\mathrm{air} \,A_\mathrm{UAV}}{2 \,m_\mathrm{UAV} \,g}$, the \gls{pdf} of $\alpha_\mathrm{UAV}$ is derived as follows:
\begin{multline}
    \label{eq:pdf_alpha}
        f_{\alpha_\mathrm{UAV}}(\alpha_\mathrm{UAV}) = f_{V^2}\!\left(\frac{\tan\alpha_\mathrm{UAV}}{K}\right) 
    \cdot \left| \frac{d}{d\alpha}\frac{\tan\alpha_\mathrm{UAV}}{K} \right|  \\
    =  
    \displaystyle\frac{\left(\tfrac{\tan\alpha_\mathrm{UAV}}{K}\right)^{\alpha_\Gamma - 1} 
    \exp\!\left(-\tfrac{\tan\alpha_\mathrm{UAV}}{K\beta_\Gamma}\right)  }
    {\Gamma(\alpha_\Gamma)\,\beta_\Gamma^{\alpha_\Gamma}}\cdot\frac{\sec^2\alpha_\mathrm{UAV}}{K}.
\end{multline}

\subsection{Induced Channel Loss}\label{subsec:uav_instability} 

The effect of changes in the \gls{rx}'s orientation can be modeled as a reduction in its effective surface area. In other words, the corresponding induced channel loss $h_\alpha$ can be expressed as the ratio between the actual and the effective PD surface areas \cite{Ijeh-JOE-2021}:
\begin{equation}
    h_\alpha = \cos\alpha_\mathrm{UAV}\prod\left(\frac{\alpha_\mathrm{UAV}}{\varphi_\mathrm{Rx}}\right),
\end{equation}
where $\varphi_\mathrm{Rx}$ denotes the \gls{rx} \gls{fov}, and $\prod(u) = 1$ if $u \leq 1$, and zero otherwise.

Based on (\ref{eq:pdf_alpha}) and following the same change-of-variable approach as in the previous subsection, we obtain the \gls{pdf} of the orientation-induced losses $f_\mathrm{h_\alpha}(h)$ as:
\begin{align}
    \begin{aligned}
        f_\mathrm{h_\alpha}(h_\alpha) & = f_{\alpha_\mathrm{UAV}}(\arccos(h_\alpha)) 
    \left|\frac{d\alpha_\mathrm{UAV}}{dh_\alpha}\right| \\
    & = f_{\alpha_\mathrm{UAV}}(\arccos(h_\alpha))\cdot\frac{1}{\sqrt{1 - h_\alpha^2}},
    \end{aligned}
\end{align}
where $|.|$ denotes the absolute value.
After some mathematical manipulations, the \gls{pdf} of $h_\alpha$ is obtained as follows:
\begin{multline}
    \label{eq:final_f_h_alpha}
    f_{h_\alpha}(h) =
    \frac{1}{K \, \Gamma(\alpha_\Gamma) \, \beta_\Gamma^{\alpha_\Gamma}} \,
    h^{-(\alpha_\Gamma+1)} \\ \times (1 - h^2)^{\tfrac{\alpha_\Gamma - 2}{2}} \,
    \exp\!\left[- \frac{\sqrt{1 - h^2}}{h \, K \, \beta_\Gamma} \right].
\end{multline}

\section{Monte-Carlo Simulations for Photon Propagation}\label{sec:Monte_Carlo}

There have been numerous research works dedicated to \gls{uwoc} channel modeling based on experimental measurements \cite{Oubei-OpticsLetters-2017, Zedini-TransComm-2019}, analytical approaches based on the \gls{rte}, e.g., \cite{Li-WCL-2015, Jaruwatanadilok-JSAC-2008}, or otherwise Monte Carlo simulations. This latter approach offers the advantage of simplicity, as it estimates the channel gain or impulse response by simulating the propagation of a large number of photons \cite{Gabriel-JOCN-2012,Liu-OptEng-2017, Enghiyad-JOSA-2021}. 

In this study, we employ the Monte Carlo method to analyze the channel effects in the \gls{w2a} transmission. For this purpose, an algorithm that effectively incorporates the influence of the underwater channel, surface waves, and air bubbles has been developed. Given the unique characteristics of the application scenario under consideration and the complexity and cost associated with the required extensive experimental measurements, this is indeed an appropriate approach for evaluating channel effects and understanding the impact of the various phenomena on the link performance.
As depicted in Fig.\,\ref{fig:model_illustration}, we consider an \gls{led} at the \gls{tx} side positioned at a depth $d_{\mathrm{water}}$ below the sea surface, and an \gls{sipm} \gls{pd} at the \gls{rx} located at a height $d_{\mathrm{air}}$ above the sea surface.
\subsection{Photon Propagation in Water}
For each photon, we initialize its position, direction, and weight. The initial position of the photons $(x_p^0, y_p^0,z_p^0)$ corresponds to the position of the \gls{tx} $(x_{\mathrm{Tx}}, y_{\mathrm{Tx}}, z_{\mathrm{Tx}})$. To initialize the photon's direction, considering the Lambertian radiation pattern for the \gls{led}, we need to initialize the zenith angle $\theta_{\mathrm{ini}}$ and the azimuthal angle $\phi_{\mathrm{ini}}$ as follows:
\begin{equation}
    \label{theta_phi_ini}
    \begin{cases}\theta_{\mathrm{ini}} = \arccos[(1 - \zeta^\theta)^{\frac{1}{1+m}}],\\ \phi_{\mathrm{ini}} = 2\pi\zeta^\phi.\end{cases}
\end{equation}
Here, $\zeta^\theta$ and $\zeta^\phi$ are independent random variables with distribution ${\cal{U}}(0,1)$, whereas $m$ represents the \gls{led} Lambertian order, given by $m = -\log2/\log\cos\theta_{1/2}$, with $\theta_{1/2}$ representing the \gls{led} half-angle \cite{Ghassemlooy-CRC-2017}. In Cartesian coordinates, the photon's initial direction vector $(\mu_x^0, \mu_y^0, \mu_z^0)$ is then given by: 
\begin{equation}
    \label{mu_zero}
    (\mu_x^0, \mu_y^0, \mu_z^0) = (\cos\phi_{\mathrm{ini}}\sin\theta_{\mathrm{ini}} ~,~ \sin\phi_{\mathrm{ini}}\sin\theta_{\mathrm{ini}} ~,~ \cos\theta_{\mathrm{ini}}).
\end{equation}

We set the initial weight of the photon $W^0$ to $1$. Then, we calculate a random step $\Delta_s$ representing the distance the photon travels until interaction with a water molecule, a particle in water, or a bubble (see Subsection\,\ref{Subsec-effect-air-bubbles}). For this, we generate a random variable $\xi$, with distribution ${\cal{U}}(0,1)$, and obtain the step as $\Delta_s = -\log\xi / c(z)$, where $c(z)$ represents the attenuation coefficient given by $c(z) = a + b + b_{\mathrm{bub}}(z)$, where $a$ denotes the absorption coefficient, and $b$ and $b_{\mathrm{bub}}(z)$ represent the water and bubble scattering coefficients, respectively (all in units of m$^{-1}$) \cite{Sahoo-Optics-2022}. At this position, we check whether the photon has reached the sea surface. If the photon is still in water, (at step $i$) we then calculate its new position (at step $i+1$) as follows:
\begin{align}
    \label{position_update}
    \begin{aligned}
        x_p^{i + 1} & = x_p^i + \Delta_s\,\mu_x^i,\\
        y_p^{i + 1} & = y_p^i + \Delta_s\,\mu_y^i,\\
        z_p^{i + 1} & = z_p^i + \Delta_s\,\mu_z^i.
    \end{aligned}
\end{align}
To check whether or not the photon has been deviated from its initial direction due to scattering, we generate a random variable $\xi_1$ with distribution ${\cal{U}}(0,1)$, and compare it with the single scattering albedo $\omega_\mathrm{al} = (b + b_{\mathrm{bub}})/c(z)$ \cite{Sahoo-Optics-2022}. Then, if $\xi_1 \leq \omega_\mathrm{al}$, we deduce that the photon has been scattered. Now, to check whether the scattering effect has been produced by a water molecule, a particle in water, or a bubble, we generate another random variable $\xi_2$ with distribution ${\cal{U}}(0,1)$, and compare it to $b_{\mathrm{bub}}/(b+b_{\mathrm{bub}})$ \cite{Sahoo-Optics-2022}. Then, (i) if $\xi_2 \leq b_{\mathrm{bub}}/(b+b_{\mathrm{bub}})$ the photon is considered to be scattered by an air bubble, in which case we calculate the new zenith angle $\theta_s$ using the Mie scattering theory; (ii) otherwise, the photon is considered to be scattered by a water molecule or a particle in water, in which case we calculate $\theta_s$ using the \gls{ff} function. 

Note, in both cases, scattering is symmetrical with respect to the azimuth, and therefore, the new azimuthal angle $\phi_s$ is generated considering the distribution ${\cal{U}}(0,2\pi)$. The elements of the new direction vector are then as follows.
\begin{align}
    \label{mu_update}
    \begin{aligned}
        & \mu_x^{i+1}=\frac{\sin \theta_s\left(\mu_x^i \mu_z^i \cos \phi_s-\mu_y^i \sin \phi_s\right)}{\sqrt{1-\left(\mu_z^i\right)^2}}+\mu_x^i \cos \theta_s, \\
        & \mu_y^{i+1}=\frac{\sin \theta_s\left(\mu_y^i \mu_z^i \cos \phi_s+\mu_x^i \sin \phi_s\right)}{\sqrt{1-\left(\mu_z^i\right)^2}}+\mu_y^i \cos \theta_s, \\
        & \mu_z^{i+1}=-\sin \theta_s \cos \phi_s \sqrt{1-\left(\mu_z^i\right)^2}+\mu_z^i \cos \theta_s .
    \end{aligned}
\end{align}
The photon weight is then updated as $W^{i+1} = W^i\,\omega_\mathrm{al}$. This operation is repeated as long as the photon remains in water. 

Algorithm~\ref{alg:scattering} specifies the pseudo-code used to model absorption and scattering in the simulation framework.

\begin{algorithm}
\caption{\textsc{HandleUnderwaterScattering}}
\label{alg:scattering}
\begin{algorithmic}[1]
\REQUIRE Current photon state $(x_p^i, y_p^i, z_p^i)$, $\vec{\mu}^i$, $W^i$, channel parameters
\ENSURE Updated position, direction $\vec{\mu}^{i+1}$, weight $W^{i+1}$, and step $i$

\STATE Update position $(x_p^{i+1}, y_p^{i+1}, z_p^{i+1})$ using \eqref{position_update}
\STATE Generate $\xi_1 \sim \mathcal{U}(0,1)$

\IF{$\xi_1 \le \omega_{al}$} 
    \STATE Generate $\xi_2 \sim \mathcal{U}(0,1)$
    \IF{$\xi_2 \le b_\mathrm{bub}/(b+b_\mathrm{bub})$}
        \STATE Calculate scattering angle $\theta_s$ using Mie theory
    \ELSE
        \STATE Calculate scattering angle $\theta_s$ using Fournier-Forand function
    \ENDIF
    \STATE Generate azimuthal scattering angle $\phi_s \sim \mathcal{U}(0,2\pi)$
    \STATE Update direction vector $\vec{\mu}^{i+1}$ using \eqref{mu_update}
\ELSE
    \STATE $\vec{\mu}^{i+1} \gets \vec{\mu}^{i}$ \COMMENT{Photon is not scattered (direction preserved)}
\ENDIF
\STATE Update photon weight $W^{i+1} \gets W^i \cdot \omega_{al}$
\STATE $i \gets i + 1$
\end{algorithmic}
\end{algorithm}

\subsection{Photon Propagation at the Water-Air Interface}
\begin{figure}
    \centering
    \includegraphics[width=0.3\textwidth]{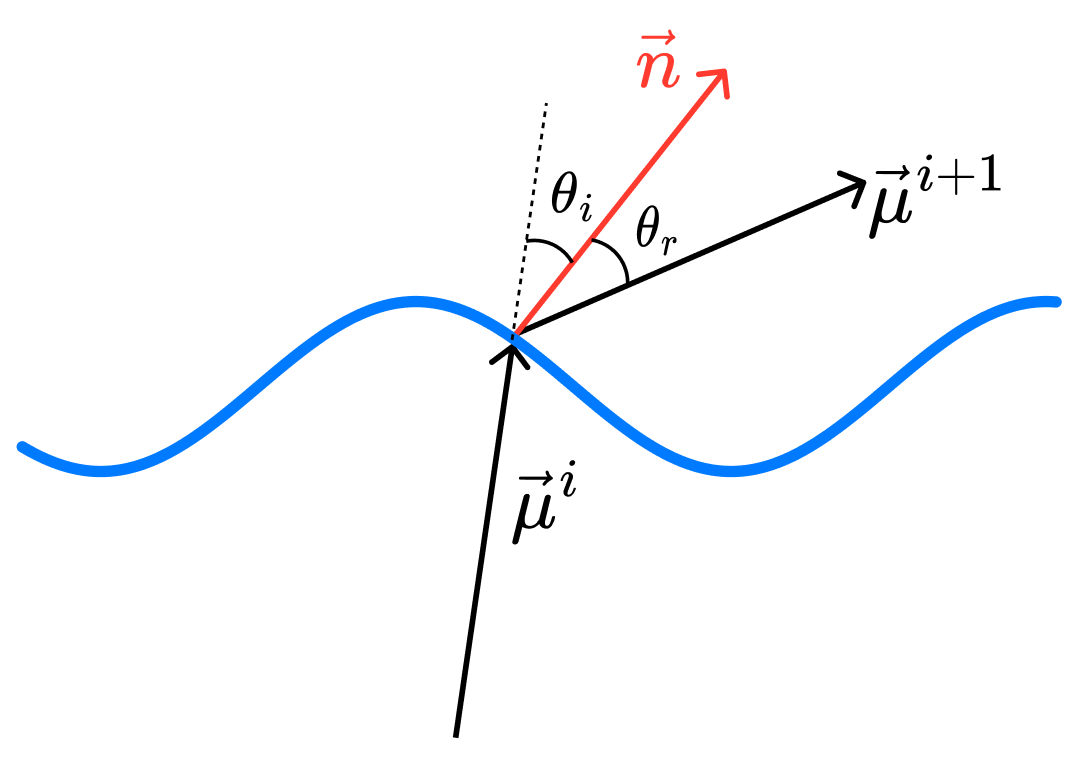}
    \caption{\small Illustration of Snell's law at photon arrival at the water-air interface.}
    \label{fig:angles_fig}
\end{figure}

As soon as the photon reaches the sea surface, we first determine the point of intersection with the sea surface $(x_w^i, y_w^i, z_w^i)$, and the vector normal to this point $\vec{n} = (\hat{n_x}, \hat{n_y}, \hat{n_z})$. We then calculate the incident angle $\theta_i$ formed between the photon direction vector $\vec{\mu}^i = (\mu_x^i, \mu_y^i, \mu_z^i)$ and the normal vector as follows \cite{Lin-JLT-2022}, see Fig.\,\ref{fig:angles_fig}:
\begin{equation}
    \label{theta_i}
    \theta_i = \arccos\left(\frac{\vec{\mu}^i.\vec{n}}{||\vec{\mu}^i||.||\vec{n}||}\right ),
\end{equation}
where $(.)$ and $||.||$ denote scalar vector product and Frobenius norm, respectively.

As illustrated in Fig. \ref{fig:angles_fig}, the Snell's law is used to obtain the refracted angle $\theta_r$ after the sea surface, before checking that $\theta_i$ is not greater than the critical angle $\theta_c = \arcsin(n_2/n_1) \approx 48. 75^o$, derived from Snell's law, with $n_1=1.33$, and $n_2=1$ the refractive indices of water and air, respectively. If $\theta_i > \theta_c$, the photon will be reflected (i.e., remain in water) and will therefore be considered as lost. To simulate photon propagation more realistically, we also account for the effect of partial reflection governed by Fresnel equations \cite{Qin-Optics-2022}. As a result, a fraction of photons with an incidence angle smaller than $\theta_c$ will be randomly reflected as well. To consider these latter, we generate a random variable $\xi_3$ with distribution ${\cal{U}}(0,1)$, and compare it with the Fresnel reflection coefficient $\rho$, given by \cite{Qin-Optics-2022}:
\begin{equation}
    \label{rho}
    \rho = \begin{cases} \frac{1}{2} \left\{ \left[\frac{\sin(\theta_i - \theta_r)}{\sin(\theta_i + \theta_r)}\right]^2 + \left[\frac{\tan(\theta_i - \theta_r)}{\tan(\theta_i + \theta_r)}\right]^2 \right\} &,\quad \mathrm{if }\, \theta_i \neq 0, \\ \left(\frac{n_1 - n_2}{n_1 + n_2}\right)^2 &,\quad \mathrm{otherwise,} \end{cases}
\end{equation}
where $\theta_r$ is calculated according to Snell's law as follows:
\begin{equation}
    \label{theta_r}
    \theta_r = \arcsin\left( \frac{n_2}{n_1}\sin\theta_i \right).
\end{equation}
Then, if $\xi_3 < \rho$ is verified, the photon will be reflected and will therefore be considered as lost \cite{Qin-Optics-2022}.  
For photons that have not been reflected, we obtain their new direction using the following equation \cite{Lin-JLT-2022}:
\begin{equation}
    \label{mu_post_surface}
    \vec{\mu}^{i+1} = \cos(\theta_r)\vec{n} + \sin\left(\theta_r\frac{\vec{\mu}^i - \vec{n}(\frac{\vec{\mu}^i.\vec{n}}{||\vec{n}||})}{\big|\big|\vec{\mu}^i - \vec{n}(\frac{\vec{\mu}^i.\vec{n}}{||\vec{n}||}) \big|\big|}\right).
\end{equation}
We also calculate the new weight $W^{i+1}$ of the photon, which we consider to be unpolarized, using the Fresnel transmission coefficient $T_f = (T_s + T_p)/2$ with $T_s$ and $T_p$ the transmission coefficients for the S (longitudinal) and P (transverse) components, respectively:
\begin{equation}
    \label{T_s_T_p}
    \begin{aligned}
        T_s & =\frac{\sin (2 \theta_i) \sin (2 \theta_r)}{\sin ^2\left(\theta_i+\theta_r\right)}, \\
        T_p & =\frac{\sin (2 \theta_i) \sin (2 \theta_r)}{\sin ^2\left(\theta_i+\theta_r\right) \cos ^2\left(\theta_i-\theta_r\right)}.
    \end{aligned}
\end{equation}
In addition, given that the sea surface is not smooth, we consider a transmission coefficient $T_u$, which depends on the wind speed $U_{10}$, as follows \cite{Qin-Optics-2022}:
\begin{equation}
    \label{T_u}
    T_{u}= \begin{cases}
    1-1.2 \times 10^{-5}\  U_{10}^{3.3}\ , & U_{10} \leq 9\ \mathrm{m}/\mathrm{s}, \\ 
    1-1.2 \times 10^{-5}\  U_{10}^{3.3} \times \\
\quad (0.255U_{10} - 0.99), & U_{10} > 9\ \mathrm{m}/\mathrm{s}.
    \end{cases}
\end{equation}

Then, the new photon weight is obtained as 
\begin{equation}\label{new-w}
W^{i+1} = W^i\,T_f\,T_u.
\end{equation}

Overall, from (\ref{position_update}), (\ref{mu_post_surface}), (\ref{T_u}), and (\ref{new-w}), we can calculate the intersection position, the direction, and the weight of the photons.
\subsection{Photon Propagation in the Air}

If a photon reaches the air, it is assumed to propagate in a straight line from that point onward, as interactions between the photon and air particles are considered to be negligible. 
Consequently, we update the position of the photon at height~$z_{\mathrm{Rx}}$. Then,  
two conditions should be simultaneously verified to determine whether the photon falls within the \gls{rx}'s \gls{fov} $\varphi_{\mathrm{Rx}}$ and active area of radius $r_{\mathrm{Rx}}$, namely:
\begin{multline}
    \label{condition_2}
    \left(x_p - x_{\mathrm{Rx}}\right)^2 + \left(y_p - y_{\mathrm{Rx}}\right)^2 < \\ \left(r_{\mathrm{Rx}} + \left(z_{\mathrm{Rx}} - z_p\right) \tan \frac{\varphi_{\mathrm{Rx}}}{2}\right)^2,
\end{multline}
and,
\begin{multline}
    \label{condition_1}
    \left(x_p + \frac{\mu_x\left(z_{\mathrm{Rx}} - z_p\right)}{\mu_z} - x_{\mathrm{Rx}}\right)^2 + \\
    \left(y_p + \frac{\mu_y\left(z_{\mathrm{Rx}} - z_p\right)}{\mu_z} - y_{\mathrm{Rx}}\right)^2 < r_{\mathrm{Rx}}^2.
\end{multline}

It is important to note that  (\ref{condition_2}) and (\ref{condition_1}) assume perfect beam alignment, i.e., no inclination with respect to the optical axis. In fact, the effect of beam misalignment (corresponding to the inclination angle $\alpha_\mathrm{UAV}$ of the UAV caused by turbulent winds) is already accounted for in $h_\alpha$, as described in Subsection~\ref{sec:uav_instability}.\ref{subsec:uav_instability}.

Algorithm~\ref{alg:interface} describes the sub-procedure used for handling complex physical interactions at the dynamic sea surface and during photon detection.

\begin{algorithm}
\caption{\textsc{HandleInterfaceCrossing}}
\label{alg:interface}
\begin{algorithmic}[1]
\REQUIRE Current photon state, interface normal vector $\vec{n}$, Rx parameters
\ENSURE Updated $W_\mathrm{total\_received}$ and $\mathrm{active}$ status

\STATE Find intersection point $(x_w^i, y_w^i, z_w^i)$
\STATE Calculate incidence angle $\theta_i$ using \eqref{theta_i}

\IF{$\theta_i \le \theta_c$}
    \STATE Calculate Fresnel reflection coefficient $\rho$ using \eqref{rho}
    \STATE Generate $\xi_3 \sim \mathcal{U}(0,1)$
    
    \IF{$\xi_3 < \rho$}
        \STATE $\mathrm{active} \gets \mathbf{false}$ \COMMENT{Photon reflected back into water (terminated)}
    \ELSE
        \STATE Calculate refraction angle $\theta_r$ using \eqref{theta_r}
        \STATE Update direction vector $\vec{\mu}^{i+1}$ using \eqref{mu_post_surface}
        \STATE Calculate transmission coefficients $T_f$ and $T_u$ using \eqref{T_s_T_p} and \eqref{T_u}
        \STATE Update photon weight $W^{i+1} \gets W^i T_f T_u$
        \STATE Propagate photon in a straight line to Rx plane at $z_\mathrm{Rx}$
        
        \IF{photon satisfies Rx active area \eqref{condition_2} and FoV \eqref{condition_1} conditions}
            \STATE $W_\mathrm{total\_received} \gets W_\mathrm{total\_received} + W^{i+1}$
        \ENDIF
        \STATE $\mathrm{active} \gets \mathbf{false}$ \COMMENT{Photon trajectory ends in air}
    \ENDIF
\ELSE
    \STATE $\mathrm{active} \gets \mathbf{false}$ \COMMENT{Total internal reflection (terminated)}
\ENDIF

\end{algorithmic}
\end{algorithm}

\subsection{Monte-Carlo Channel Loss}
\begin{figure*}[ht]
    \centering
    \includegraphics[width=0.98\textwidth]{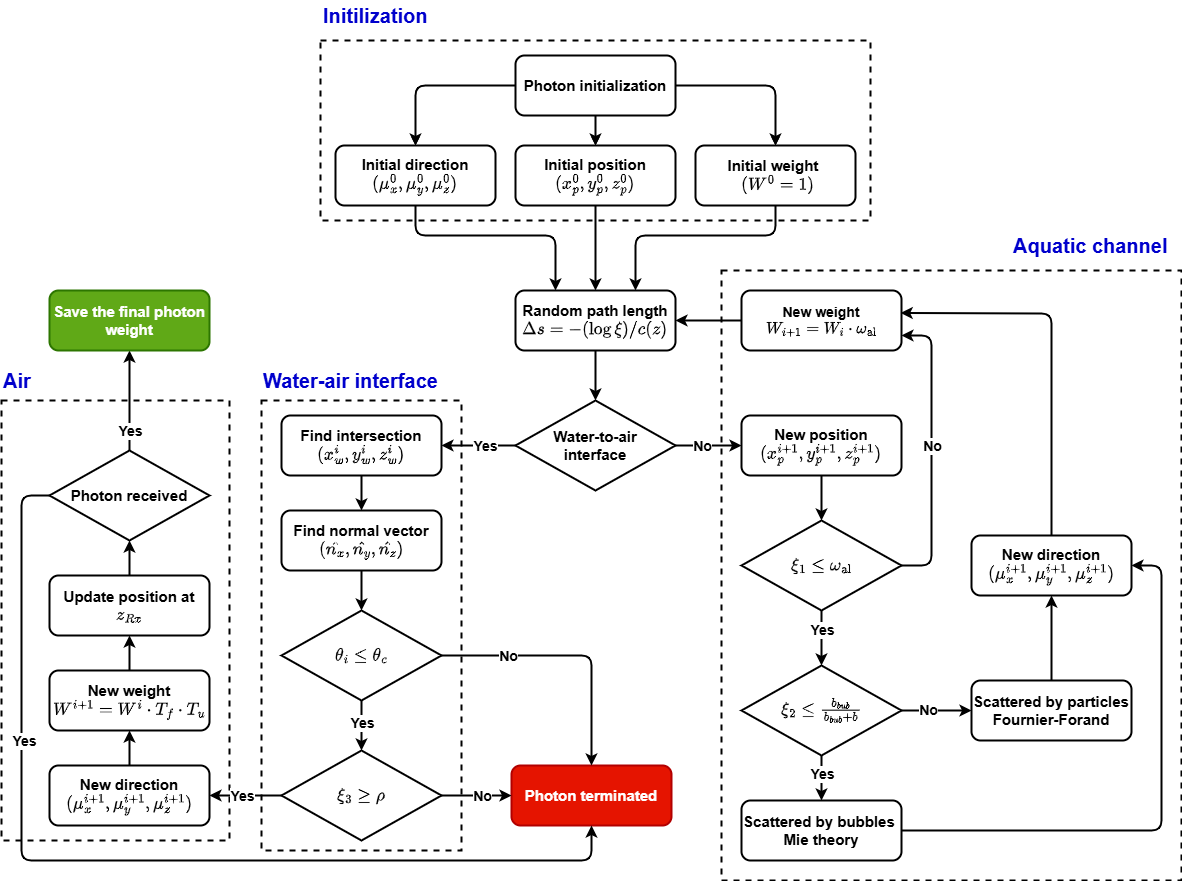}
    \caption{\small Flow chart of Monte Carlo photon propagation algorithm for the W2A link.}
    \label{fig:monte_carlo_scheme}
\end{figure*}

Finally, the channel loss $h_\mathrm{MC}$ is obtained by calculating the ratio of the sum of the weights of the received photons to the total number of emitted photons in the simulation. Figure \ref{fig:monte_carlo_scheme} shows the overall methodology, while Algorithm~\ref{alg:MC_main} presents the pseudo-code for the main ray-tracing engine. This primary algorithm orchestrates the overall photon tracking loop by invoking the previously defined sub-procedures, i.e., Algorithm~\ref{alg:scattering} for the underwater propagation phase, followed by Algorithm~\ref{alg:interface} to resolve the surface boundary interactions.

\begin{algorithm}
\caption{Main Monte Carlo Ray-Tracing Engine for W2A Channel}
\label{alg:MC_main}
\begin{algorithmic}[1]
\REQUIRE $N_\mathrm{photons}$: Total number of transmitted photons
\REQUIRE Tx, Rx, and Channel parameters
\ENSURE Monte-Carlo channel loss $h_\mathrm{MC}$

\STATE $W_\mathrm{total\_received} \gets 0$

\FOR{$k = 1$ to $N_\mathrm{photons}$}
    \STATE \textbf{Initialization:}
    \STATE Position $(x_p^0, y_p^0, z_p^0) \gets (x_\mathrm{Tx}, y_\mathrm{Tx}, z_\mathrm{Tx})$
    \STATE Generate $\zeta^\theta, \zeta^\phi \sim \mathcal{U}(0,1)$, calculate $\theta_\mathrm{ini}$ and $\phi_\mathrm{ini}$ using \eqref{theta_phi_ini}
    \STATE Set initial direction vector $\vec{\mu}^0$ using \eqref{mu_zero}
    \STATE $W^0 \gets 1$, step index $i \gets 0$, $\mathrm{active} \gets \mathbf{true}$
    
    \WHILE{$\mathrm{active}$}
        \STATE Generate $\xi \sim \mathcal{U}(0,1)$
        \STATE Calculate random path length $\Delta_s \gets -\log\xi / c(z)$
        
        \IF{the photon path crosses the water-air interface}
            \STATE \textbf{Execute} \textsc{HandleInterfaceCrossing} (Algorithm~\ref{alg:interface})
        \ELSE 
            \STATE \textbf{Execute} \textsc{HandleUnderwaterScattering} (Algorithm~\ref{alg:scattering})
        \ENDIF
    \ENDWHILE
\ENDFOR

\RETURN $h_\mathrm{MC} = W_\mathrm{total\_received} / N_\mathrm{photons}$

\end{algorithmic}
\end{algorithm}
\section{Numerical Results}\label{sec:Results}
This section presents the numerical results concerning the \gls{w2a} channel characterization and link performance evaluation.
First, we illustrate the effects of sea surface waves and air bubbles on the channel gain for different \gls{tx}/\gls{rx} parameters including the consideration of link misalignment. Then, we evaluate the \gls{e2e} link performance in terms of the average \gls{ber} as a function of the \gls{led} depth for different data rates.

\subsection{System/Simulation Parameters and Main Assumptions}

Unless otherwise specified, we use the parameters shown in Table \ref{tab:link_parameters} for the \gls{w2a} link.
\begin{table}
    \centering
    \caption{Default \gls{w2a} link parameters considered in the numerical results.}
    \begin{tabular}{lcc}
        \hline
        Parameters & Symbols & Values \\
        \hline
        Absorption coefficient & $a$ & $0.114$\,m$^{-1}$ \\
         Scattering coefficient & $b$ & $0.037$\,m$^{-1}$\\
         Water refractive index & $n_1$ & $1.33$\\
         Air refractive index & $n_2$ & $1$\\
         \gls{led} wavelength & $\lambda$ & $470~\mathrm{nm}$  \\ 
         \gls{led} semi-angle at half-power & $\theta_{1/2}$ & $10^o$\\
         \gls{tx} position & $(x_{\mathrm{Tx}},y_{\mathrm{Tx}}, z_{\mathrm{Tx}})$ & $(0,0, -10)~\mathrm{m}$\\
         \gls{rx} position & $(x_{\mathrm{Rx}},y_{\mathrm{Rx}}, z_{\mathrm{Rx}})$ & $(0,0, 5)~\mathrm{m}$\\
         \gls{rx} \gls{fov} & $\varphi_{\mathrm{Rx}}$ & $60^o$\\
         \gls{rx} active area radius & $r_{\mathrm{Rx}}$ & $5~\mathrm{cm}$\\
         \hline
    \end{tabular}
    \label{tab:link_parameters}
\end{table}
The considered values for the absorption and scattering coefficients $a$ and $b$ correspond to the case of clear ocean waters \cite{Mobley-CRC-1994}. Indeed, the validity of this assumption can be verified, for example, in \cite{Blondeau-JGRO-2009}, which reports the values of absorption and scattering coefficients observed in Australian reef waters.
Additionally, as previously established in Subsection\,\ref{Subsec:Model-water} for the photon scattering \gls{ff} \gls{spf} model, we set the refractive index to $n_s = 1.1$ and the slope parameter to $\mu_\mathrm{FF} = 3.5835$ in \eqref{FF_scat_phase_function}. As a reminder, these values correspond to a back-scattering coefficient of $b_p = 0.0183$\,m$^{-1}$, representative of clear ocean waters \cite{Mobley-AOptics-2002}. Furthermore, the fetch parameter $F$ is set to $30~\mathrm{km}$.
The \gls{tx} and the \gls{rx} are assumed to be perfectly aligned, and data transfer is carried out at night in order not to be affected by background radiations, given the highly sensitive \gls{sipm} employed at the \gls{rx}. Given the relatively small active area of the \gls{sipm} (see Table\,\ref{tab:sipm_parameters}), a non-imaging lens is considered in front of the \gls{pd} to expand the active area of the \gls{rx} without compromising its \gls{fov}. Here, the considered radius for the \gls{rx} active area is $5$~cm.

To study the sea surface effect, we consider two different wind speeds\footnote{The \gls{jonswap} model is valid for $U_{10}$ between $0$ and $20$\,m/s.} of $U_{10} = 5$, and $13$\,m/s. The case of $U_{10} = 0$, representing a flat sea surface with no air bubble effects, will serve as the benchmark for comparison. In the considered \gls{jonswap} model, the values of $\omega_i$ are taken within the interval $[\omega_l, \omega_h]$ with a step size of $\Delta\omega$, and the values $\theta_j$ are taken within the interval $[-\pi/2, \pi/2]$ with a step size of $\Delta\theta$, see (\ref{water_surface}) in Subsection\,\ref{Subsec-W2A-Interface}. These parameters are listed in Table\,\ref{tab:JONSWAP_parameters}, where the considered parameters in the four rightmost columns are taken from \cite{Dong-Sensors-2020}.

Finally, we consider in this study the DJI Matrice 300 RTK \gls{uav} platform~\cite{M300_RTK_AUV}. This platform was selected because it is widely employed in both industrial missions and academic research, owing to its long endurance, robustness, and overall reliability. Furthermore, it is designed to withstand wind speeds of up to $15$\,m/s, making it particularly suitable for maritime and coastal operations where gusts and turbulence are common. According to the technical specifications provided in~\cite{M300_RTK_AUV}, the \gls{uav} has a total mass of $m_\mathrm{UAV} = 8.37$\,kg (including batteries) and physical dimensions of $81 \times 67 \times 43$\,cm$^3$ ($L \times W \times H$). Since the values of $A_\mathrm{UAV}$ and $C_D$ (see equation~(\ref{eq:tan_alpha})) are not publicly available, we estimated them using ANSYS software, obtaining $0.250\,\text{m}^2$ and $0.8$, respectively.

For the sake of self-containment, it is important to clearly define the applicability range of the physical assumptions made in the proposed channel model. First, in this work, we exclusively consider ``air'' bubbles (i.e., without coating of organic film~\cite{OceanOptics_Website}), which allows us to neglect their absorption effects, as they are orders of magnitude weaker than their scattering effects \cite{OceanOptics_Website}. Furthermore, the effect of oceanic turbulence is neglected due to the focus on shallow-water environments, such as coral reef regions. This assumption is corroborated using real-world \gls{aodn} measurement data in Appendix~\ref{app:temp_sal}, allowing us to focus primarily on the phenomena specific to \gls{w2a} transmissions. Finally, the considered wind speed range is consistent with the operational limits of the considered \gls{uav} platform, which is widely deployed in maritime missions.
\begin{table}
    \caption{Angular frequency range (in rad/s) and directional angles (in rad) used for sea surface simulation for the three considered wind speeds $U_{10}$ (in m/s).}
    \centering
    \begin{tabular}{clcccc}
        \hline
         $U_{10}$ & $[\omega_l, \omega_h]$ & $\Delta\omega$ & $M$ & $\Delta\theta$ & $N$\\
         \hline
            5 & $[1.2, 4.0]$ & 0.2 & 15 & $\pi/5$ & 6\\
            13 & $[0.9, 3.0]$ & 0.1 & 22 & $\pi/20$ & 21\\
         \hline
    \end{tabular}
    \label{tab:JONSWAP_parameters}
\end{table}

\subsection{Effect of UAV Deflection Angle}\label{subsec:effect_uav_angle}

\begin{figure}
    \centering
    \subfloat[]{\includegraphics[width=0.45\textwidth]{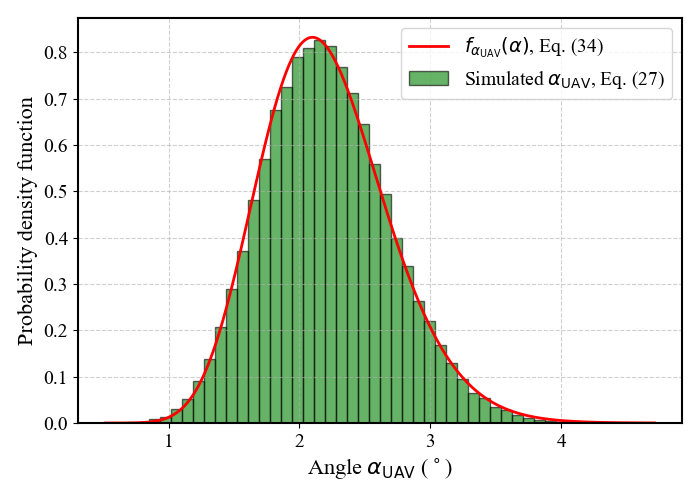}%
    \label{angle_u_5}}\\
    \subfloat[]{\includegraphics[width=0.45\textwidth]{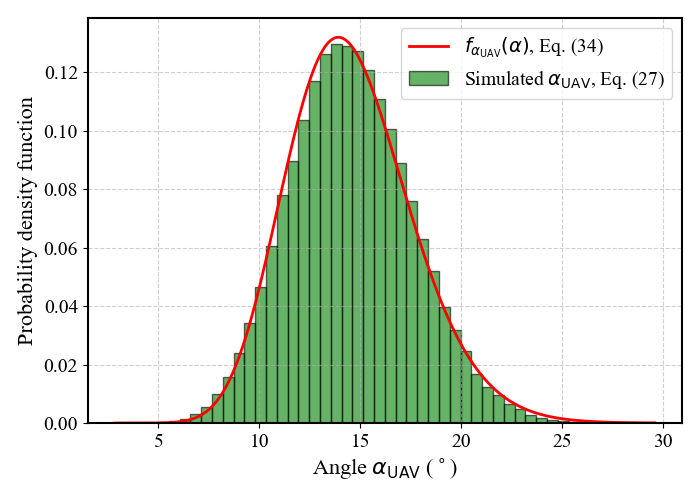}%
    \label{angle_u_13}}
    \caption{\small Simulation-based distribution of $\alpha_\mathrm{UAV}$ using (\ref{eq:tan_alpha}) and its analytical distribution using (\ref{eq:pdf_alpha}) for wind speeds: (a) $U_{10}=5$~m/s, and (b) $U_{10}=13$~m/s}.
    \label{fig_angles}
\end{figure}

To investigate the effect of \gls{uav} instability, we first analyze the distribution of $\alpha_\mathrm{UAV}$, in order to validate the accuracy of the proposed approximate analytical \gls{pdf}, $f_{\alpha_\mathrm{UAV}}(\alpha)$. 
Figures~\ref{angle_u_5} and \ref{angle_u_13} contrast the simulation-based histograms of $\alpha_\mathrm{UAV}$ obtained from (\ref{eq:tan_alpha}), with $f_{\alpha_\mathrm{UAV}}(\alpha)$ given by the closed-form expression in (\ref{eq:pdf_alpha}), for the two cases of $U_{10} = 5$ and $13$~m/s, respectively.

First, we observe that the proposed closed-form expression for the \gls{pdf} provides a highly accurate approximation of the simulation-based distribution of $\alpha_{\mathrm{UAV}}$. To quantitatively validate the goodness of fit, we evaluate the \gls{rmse}, \gls{mae},  Kolmogorov-Smirnov statistic $D_{\mathrm{KS}}$, and additionally, the Kullback-Leibler divergence $D_{\mathrm{KL}}$ between the analytical and empirical distributions \cite{Wasserman-book-2004}. For $U_{10} = 5$\,m/s, we obtain an \gls{rmse} of $1.12 \times 10^{-2}$, a \gls{mae} of $7.71 \times 10^{-3}$, a $D_{\mathrm{KS}}$ of $2.95 \times 10^{-2}$, and a $D_{\mathrm{KL}}$ of $1.45 \times 10^{-3}$. Similarly, for $U_{10} = 13$\,m/s, the corresponding metrics yield an \gls{rmse} of $2.5 \times 10^{-3}$, a \gls{mae} of $1.74 \times 10^{-3}$, a $D_{\mathrm{KS}}$ of $6.9 \times 10^{-3}$, and a $D_{\mathrm{KL}}$ of $3.13 \times 10^{-3}$. These consistently low error metrics confirm that the proposed Gamma approximation provides a mathematically tractable and physically robust framework across different turbulence conditions. Second, we note the influence of increasing $U_{10}$ on the \gls{uav}'s inclination angle: as $U_{10}$ increases, larger tilt angles with a higher variance are induced. Specifically, the standard deviation of $\alpha_{\mathrm{UAV}}$ increases from $\sigma_{\alpha} = 0.479^{\circ}$ at $U_{10} = 5$\,m/s to $\sigma_{\alpha} = 3.04^{\circ}$ at $U_{10} = 13$\,m/s, indicating reduced stability and greater variability under stronger wind turbulence. Nevertheless, given the selected \gls{uav} (a relatively heavy platform designed to operate under strong wind conditions), the inclination angles remain bounded. This observation is further confirmed in the analysis of $h_{\alpha}$, described below.

\begin{figure}
    \centering
    \subfloat[]{\includegraphics[width=0.45\textwidth]{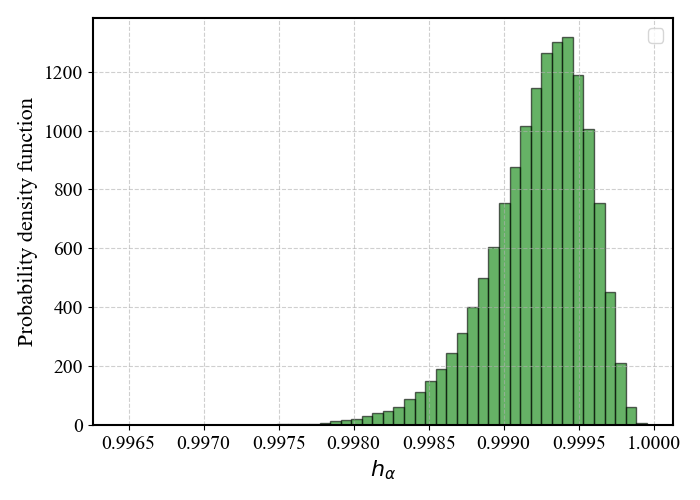}%
    \label{angle_h_alpha_5}}\\
    \subfloat[]{\includegraphics[width=0.45\textwidth]{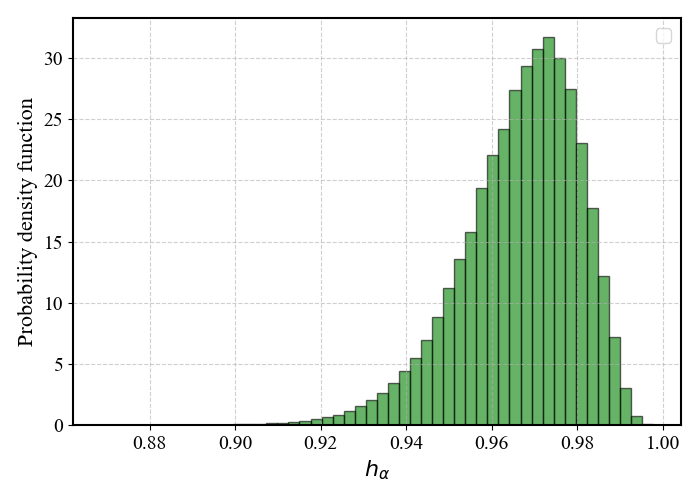}%
    \label{angle_h_alpha_13}}
    \caption{\small Distribution of $h_\alpha$ using (\ref{eq:final_f_h_alpha}) for wind speeds: (a) $U_{10}=5$~m/s, and (b) $U_{10}=13$~m/s}.
    \label{fig_h_alpha}
\end{figure}

Figures~\ref{angle_h_alpha_5} and \ref{angle_h_alpha_13} illustrate the distributions of $h_\alpha$ for $U_{10} = 5$ and $13$~m/s, respectively, which are directly governed by the previously discussed distributions of $\alpha_\mathrm{UAV}$. These results demonstrate that, owing to the selected \gls{uav} platform, the losses induced by instability remain largely negligible. It is important to emphasize that this outcome is primarily due to the use of an \gls{led} at the \gls{tx}. Indeed, the wider beam profile of the \gls{led} mitigates the impact of \gls{uav} instability, confirming the rationale behind this choice in the proposed system design.
Employing a \gls{ld}, characterized by a much narrower beam and higher sensitivity to angular deviations, would require more appropriate modeling approaches \cite{Ijeh-JOCN-2022}. In such cases, the observed inclination angles could result in significant performance degradation. A detailed investigation of this effect, considering a \gls{ld}-based \gls{tx} and quantifying the associated losses, is planned for future work.

\subsection{Assessment of the Effect of Wave Surface and Air Bubbles}

\begin{figure}
    \centering
    \subfloat[]{\includegraphics[width=0.485\textwidth]{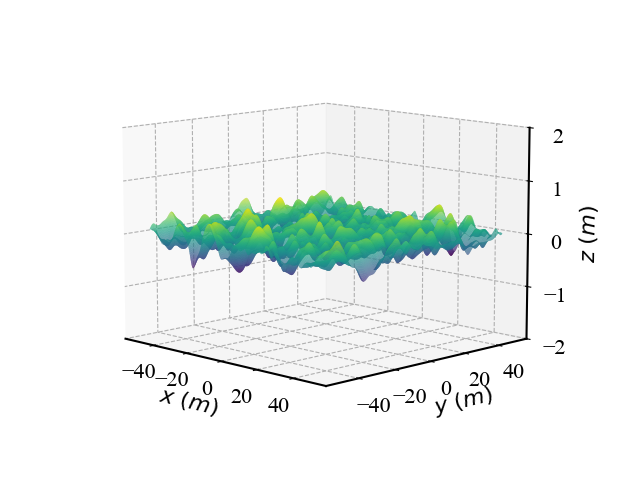}%
    \label{surface_u_5}}\\
    \subfloat[]{\includegraphics[width=0.485\textwidth]{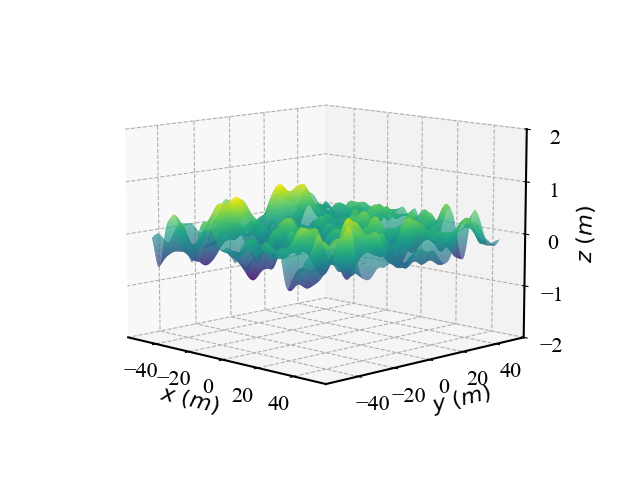}%
    \label{surface_u_13}}
    \caption{\small Simulated sea surfaces using the \gls{jonswap} wave spectrum model for the wind speeds: (a) $U_{10}=5~m/s$, and (b) $U_{10}=13~m/s$.}
    \label{fig_surface}
\end{figure}

First, based on the \gls{jonswap} model, we have presented the sea surface profiles on a $100\times100~\text{m}^2$ grid\footnote{As it can be seen on the figure, the range of considered $x$ and $y$ grids is $(-50, 50)$\,m.} for the two considered $U_{10}$ wind speeds of $5$, and $13$ m/s, in Figs.\,\ref{surface_u_5}, and \ref{surface_u_13}, respectively. 
We can observe the random nature of wave motion, along with the effect of increasing wind speed on both wave wavelength and peak-to-peak height. For instance, this latter increases from about $50~\text{cm}$ for $U_{10}=5~\text{m/s}$ to about $1.5~\text{m}$ for $U_{10}=13~\text{m/s}$.

\begin{figure}[!t]
    \centering
    \includegraphics[width=0.45\textwidth]{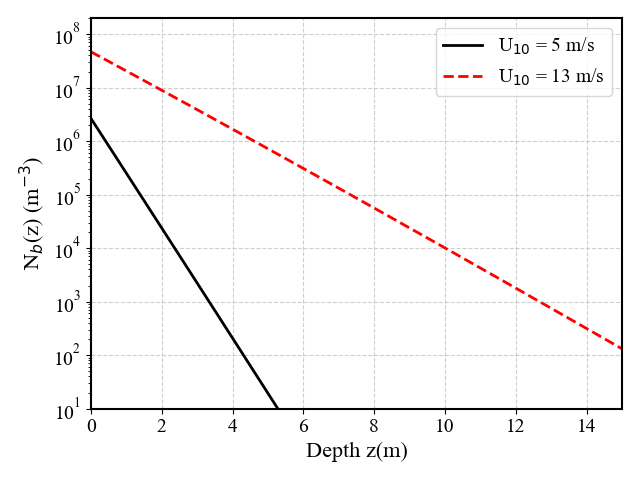}
    \caption{\small Evolution of the density number of bubbles $N_b(z)$ with depth $z$ based on the \gls{hn} bubble population model for different wind speeds $U_{10}$.}
    \label{fig_HN_density_number}
\end{figure}

\begin{figure}[!t]
    \centering
    \includegraphics[width=0.45\textwidth]{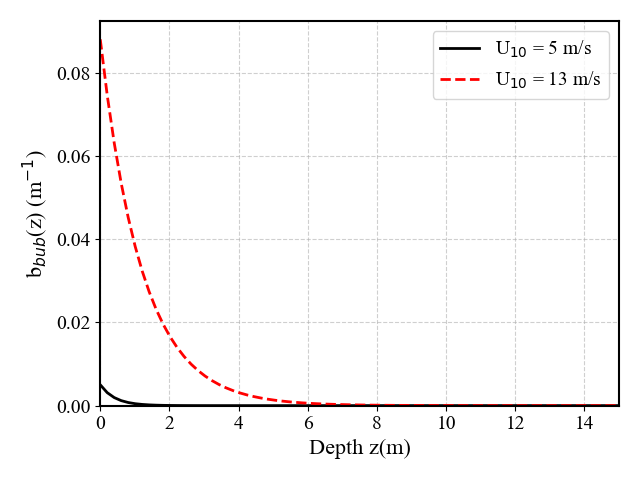}
    \caption{\small Variations of the scattering coefficient of bubbles $b_{\mathrm{bub}}$ with depth $z$ based on the HN bubble population model for different wind speeds $U_{10}$.}
    \label{fig_HN_b_bub}
\end{figure}

As concerns air bubbles effect, using the \gls{hn} model, we have presented in Fig.\,\ref{fig_HN_density_number} the evolution of $N_b(z)$ as a function of depth $z$ for the two considered wind speeds $U_{10}$. Based on this, we have presented in Fig.\,\ref{fig_HN_b_bub} the evolution of $b_{\text{bub}}$ as a function of $z$, which clearly demonstrates the significant variation in $b_{\text{bub}}$ near the surface. This highlights the importance of accounting for the effect of air bubbles on photon propagation near the water surface. Also, reasonably, higher wind speeds result in a greater concentration of air bubbles near the sea surface.
\subsection{Impact of Sea Surface Wind Speed}

\begin{figure*}[!t]
    \centering
    \subfloat[]{\includegraphics[width=0.58\textwidth]{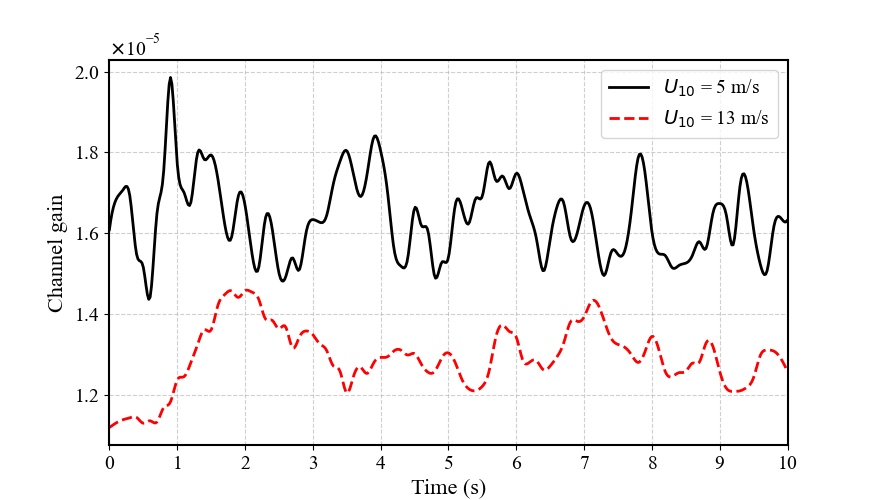}%
    \label{fig:JONSWAP_time_evolution}}
    \hfil
    \subfloat[]{\includegraphics[width=0.419\textwidth]{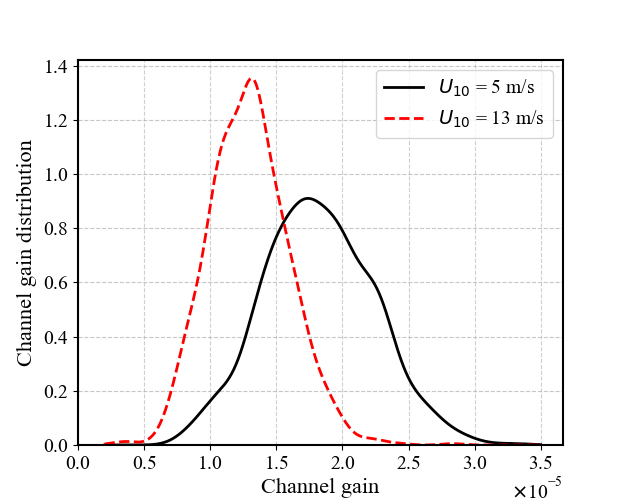}%
    \label{fig:JONSWAP_density}}
    \caption{\small W2A channel gain for different wind speeds $U_{10}$; (a) evolution over a time interval of $10$\,s; (b) distribution over $10^4$ channel realizations.}
    \label{fig:JONSWAP}
\end{figure*}

To study the effect of wind speed on the optical link performance, Fig.\,\ref{fig:JONSWAP_time_evolution} shows an example of the channel gain variations over a $10$\,s time interval for the two wind speeds under consideration. We have further shown the distribution of the channel gain for $10000$ channel realizations in Fig.\,\ref{fig:JONSWAP_density}. These results take into account the dynamics of wave motion, the presence of bubbles generated by wave breaking, as well as the \gls{uav} deflection angle due to turbulent wind speed (which can be considered negligible in this study as indicated in Subsection~\ref{sec:Results}.\ref{subsec:effect_uav_angle}).
They clearly highlight the time-varying fading effect introduced to the channel. 
As a matter of fact, increased wind speed has three notable effects. First, it alters the height and wavelength of surface waves, as illustrated in Fig.\,\ref{fig_surface}, leading to a higher photon loss during propagation from the \gls{tx} to the \gls{rx} due to the random effects of refraction and reflection at the water-air interface. Second, higher wind speeds increase the bubble population near the sea surface, resulting in a more pronounced scattering effect, as depicted in Fig.\,\ref{fig_HN_b_bub}. Finally, the effect of wind on the \gls{uav}, as described in Subsection~\ref{sec:Results}.\ref{subsec:effect_uav_angle}, increases the variation of the angle $\alpha_\mathrm{UAV}$ with increased wind intensity, thereby inducing higher losses.

For instance, from Fig.\,\ref{fig:JONSWAP_density} we can see the destructive effect of increasing wind speed on the average \gls{w2a} channel gain, which is about $1.58\times10^{-5}$, and $1.32\times10^{-5}$, for $U=5$, and $13~\text{m/s}$, respectively.

Interestingly, more significant channel gain variations are observed at lower wind speeds. This can be explained by the shorter wave periods at the surface for low $U_{10}$ values, as can be seen from Fig.\,\ref{fig_surface}. This increased variability is consistent with the findings reported in \cite[Fig.~6]{Lin-ICCC-2020}, where a greater temporal fluctuation of the channel gain was observed for $U_{10}=5$~m/s compared to $U_{10}=10$~m/s. Similar results were also presented in \cite[Fig.~13]{Lin-JLT-2022}, highlighting enhanced variability associated with small-scale surface waves.

\subsection{Effect of Tx Parameters}
\begin{figure}
\centering
\includegraphics[width=0.485\textwidth]{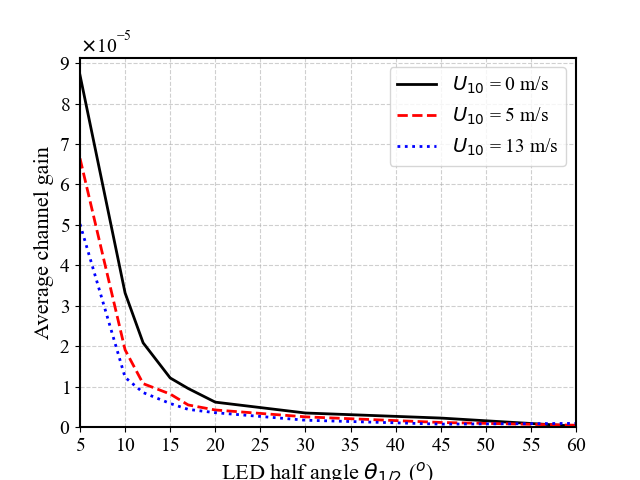}
\caption{\small Average channel gain as a function of the \gls{led} beam divergence $\theta_{1/2}$ for different wind speeds $U$.}
\label{Fig:gain_apertures}
\end{figure}

To investigate the effect of the \gls{led} beam divergence, we have shown in Fig.\,\ref{Fig:gain_apertures} the evolution of the average channel gain as a function of the \gls{led} semi-angle at half-power, $\theta_{1/2}$. Note that, in practice, $\theta_{1/2}$ can be adjusted using appropriate optics \cite{Zayed-Springer-2024}. As expected, these results highlight the significant impact of the \gls{led} beam divergence on the average channel gain. For instance, for $U_{10}=5\text{ m/s}$, increasing $\theta_{1/2}$ from $5^\circ$ to $60^\circ$ leads to a decrease in average channel gain from $6.8\times 10^{-5}$ to $2.6\times 10^{-6}$, corresponding to a loss of more than $14~\text{dB}$. A similar trend is observed for $U_{10}=13\,\text{ m/s}$.


\begin{figure*}
    \centering
    \subfloat[]{\includegraphics[width=0.3\textwidth]{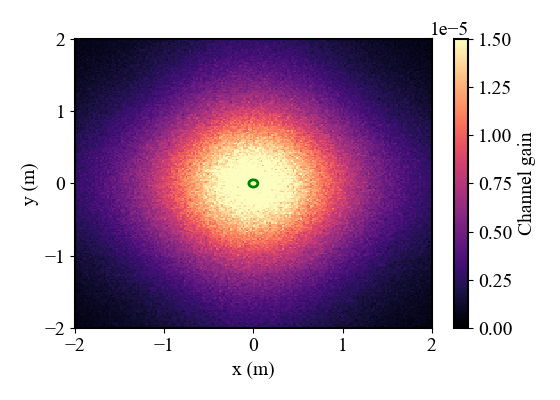}%
    \label{fig:cov_aperture_10_ws_5}}
    \hfil
    \subfloat[]{\includegraphics[width=0.3\textwidth]{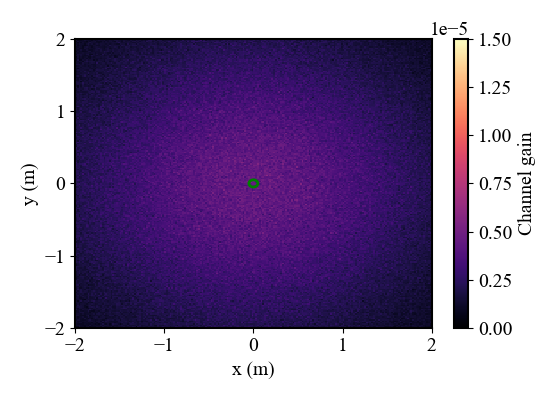}%
    \label{fig:cov_aperture_20_ws_5}}
    \hfil
    \subfloat[]{\includegraphics[width=0.3\textwidth]{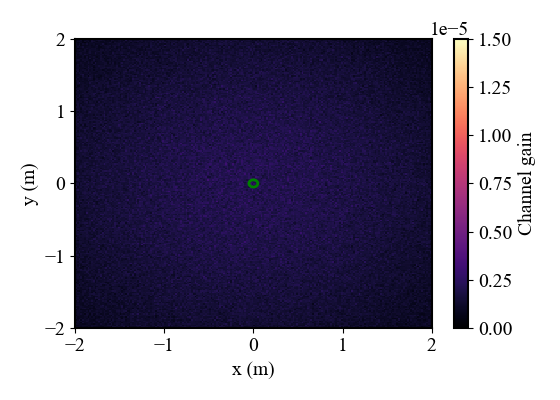}%
    \label{fig:cov_aperture_30_ws_5}}
    \hfil
    \subfloat[]{\includegraphics[width=0.3\textwidth]{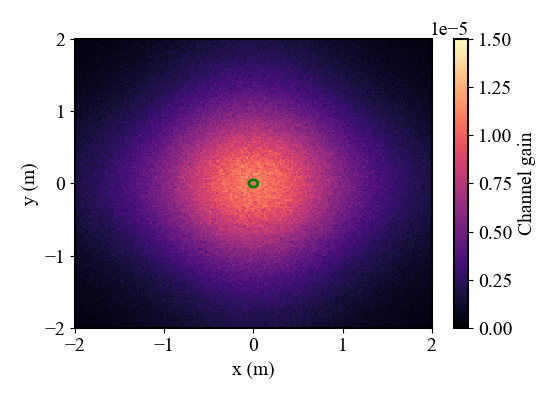}%
    \label{fig:cov_aperture_10_ws_13}}
    \hfil
    \subfloat[]{\includegraphics[width=0.3\textwidth]{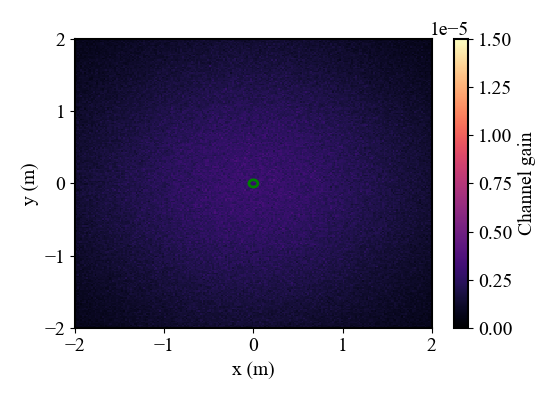}%
    \label{fig:cov_aperture_20_ws_13}}
    \hfil
    \subfloat[]{\includegraphics[width=0.3\textwidth]{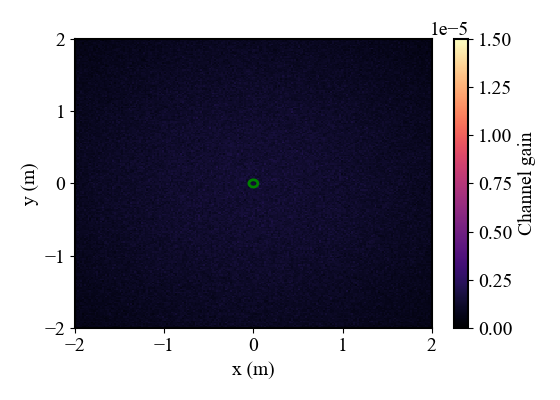}%
    \label{fig:cov_aperture_30_ws_13}}
    \caption{\small Spatial distribution of the average channel gains with different \gls{led} beam divergence angles $\theta_{1/2}$ and wind speeds $U_{10}$; (a) $U_{10}=5~\mathrm{m/s}$, $\theta_{1/2} = 10^\circ$; (b) $U_{10}=5~\mathrm{m/s}$, $\theta_{1/2} = 20^\circ$; (c) $U_{10}=5~\mathrm{m/s}$, $\theta_{1/2} = 30^\circ$, (d) $U_{10}=13~\mathrm{m/s}$, $\theta_{1/2} = 10^\circ$, (e) $U_{10}=13~\mathrm{m/s}$, $\theta_{1/2} = 20^\circ$; (f) $U_{10}=13~\mathrm{m/s}$, $\theta_{1/2} = 30^\circ$. The green circle in the center of all subplots represents the \gls{pd} active area.}
    \label{fig:coverage_aperture}
\end{figure*}

To better understand these results, we have shown in Fig.\,\ref{fig:coverage_aperture} the spatial distribution of the channel gain at the \gls{rx} plane (i.e., at $z = z_{\text{Rx}}$). More precisely, Figs.\,\ref{fig:cov_aperture_10_ws_5}, \ref{fig:cov_aperture_20_ws_5}, and \ref{fig:cov_aperture_30_ws_5} show the distributions for \mbox{$U_{10}=5~\text{m/s}$} and $\theta_{1/2} = 10^\circ$, $20^\circ$, and $30^\circ$, respectively. Similarly, Figs.\,\ref{fig:cov_aperture_10_ws_13}, \ref{fig:cov_aperture_20_ws_13}, and \ref{fig:cov_aperture_30_ws_13} show the distributions for \mbox{$U_{10}=13~\text{m/s}$}. The green circle in the figures represents the equivalent active area of the \gls{rx} according to the assumptions in Table\,\ref{tab:link_parameters}. Based on the results illustrated in these figures, it can be observed that a higher beam divergence increases photon scattering outside the \gls{rx}'s active area. This, in turn, leads to greater beam refraction at the water-air interface, resulting in a higher channel loss. Note, although using a more directional beam is more advantageous, it also increases the link sensitivity to beam misalignment (see Subsection~\ref{sec:Results}.\ref{Subsec:misalignement}).

\begin{figure}
\centering
\includegraphics[width=0.485\textwidth]{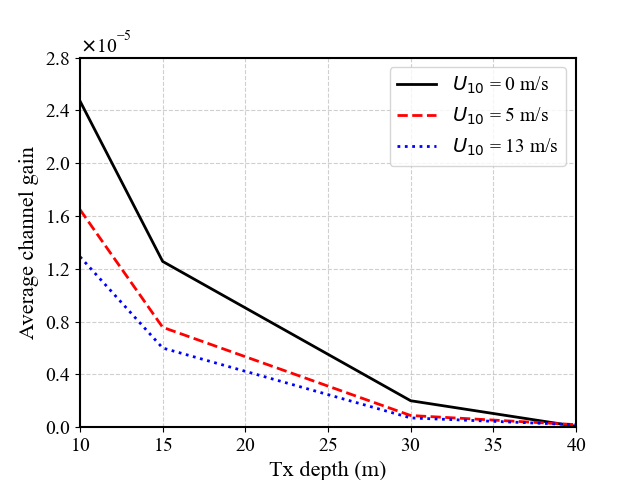}
\caption{\small Average channel gain as a function of the Tx depth $d_{\text{water}}$, $\theta_{1/2}=10^\circ$ and different wind speeds $U_{10}$.}
\label{Fig:gain_depths}
\end{figure}

To study the impact of the \gls{tx} depth, we have shown in Fig.\,\ref{Fig:gain_depths} the average channel gain as a function of $d_{\text{water}}$ varying from $10$ to $40~\text{m}$. 
A larger $d_{\text{water}}$ leads to increased photon absorption and scattering during propagation in water, and furthermore, results in a larger beam size at the water-air interface, increasing the proportion of photons not captured on the \gls{rx}'s equivalent active area.

\begin{figure*}
    \centering
    \subfloat[]{\includegraphics[width=0.3\textwidth]{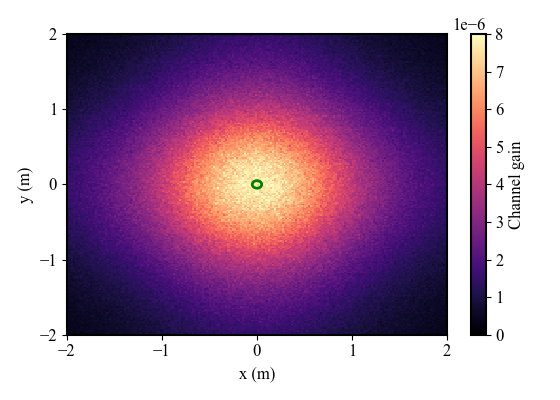}%
    \label{fig:cov_depth_15_ws_5}}
    \hfil
    \subfloat[]{\includegraphics[width=0.3\textwidth]{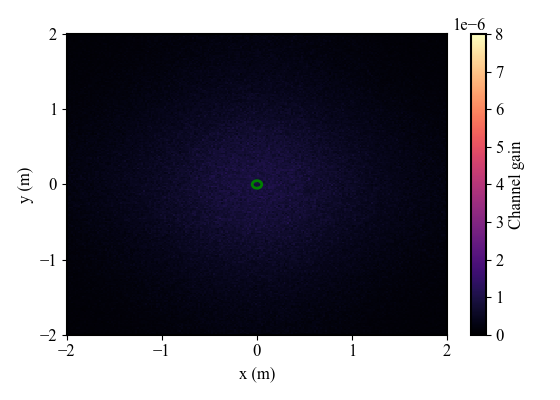}%
    \label{fig:cov_depth_30_ws_5}}
    \hfil
    \subfloat[]{\includegraphics[width=0.3\textwidth]{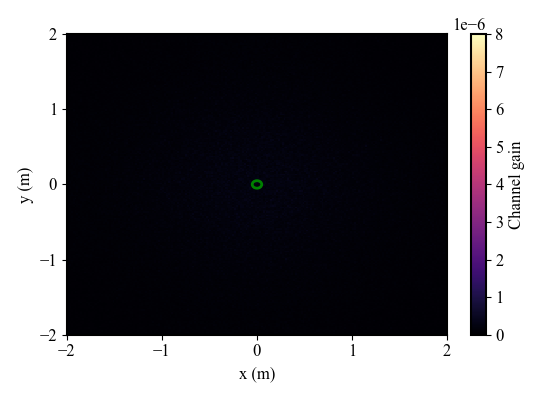}%
    \label{fig:cov_depth_40_ws_5}}
    \hfil
    \subfloat[]{\includegraphics[width=0.3\textwidth]{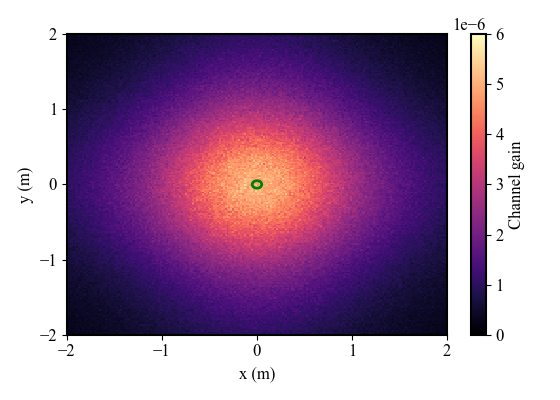}%
    \label{fig:cov_depth_15_ws_13}}
    \hfil
    \subfloat[]{\includegraphics[width=0.3\textwidth]{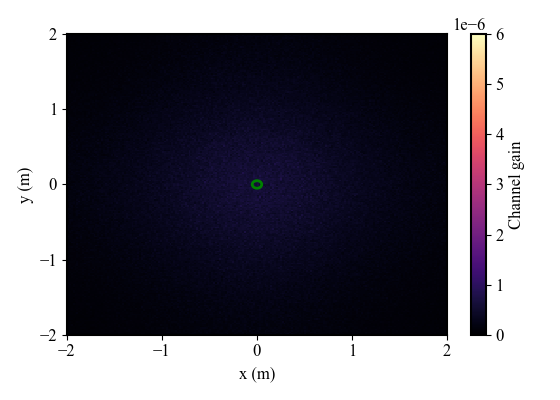}%
    \label{fig:cov_depth_30_ws_13}}
    \hfil
    \subfloat[]{\includegraphics[width=0.3\textwidth]{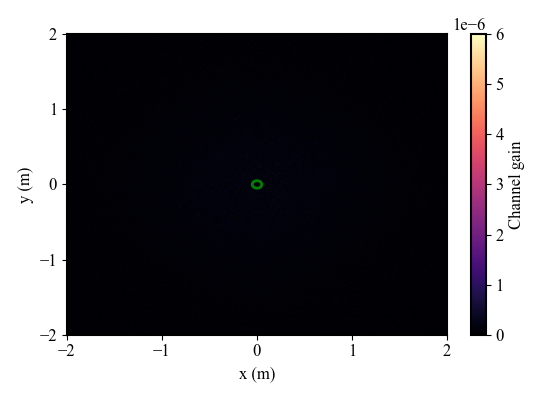}%
    \label{fig:cov_depth_40_ws_13}}
    \caption{\small Spatial distribution of the average channel gain at the \gls{rx} plane for different wind speeds and \gls{tx} depths: (a) $U_{10}=5~\mathrm{m/s}$, $d_{\mathrm{water}} = 15~\mathrm{m}$; (b) $U_{10}=5~\mathrm{m/s}$, $d_{\mathrm{water}} = 30~\mathrm{m}$; (c) $U_{10}=5~\mathrm{m/s}$, $d_{\mathrm{water}} = 40~\mathrm{m}$; (d) $U_{10}=13~\mathrm{m/s}$, $d_{\mathrm{water}} = 15~\mathrm{m}$; (e) $U_{10}=13~\mathrm{m/s}$, $d_{\mathrm{water}} = 30~\mathrm{m}$; (f) $U_{10}=13~\mathrm{m/s}$, $d_{\mathrm{water}} = 40~\mathrm{m}$. $\theta_{1/2} =10^\circ$, the green circle in the center of all subplots represents the equivalent \gls{rx} active area.}
    \label{fig:coverage_depth}
\end{figure*}

To support these results, Fig.\,\ref{fig:coverage_depth} shows the spatial distribution of the channel gain at $z=z_{\text{Rx}}$ for $d_{\text{water}}=15$, $30$, and $40~\text{m}$, and wind speeds $U_{10}=5$\,,\,$13$\,m$/$s, while $\theta_{1/2}$ is set to its default value of $10^\circ$. Compared to the results of Fig.\,\ref{fig:coverage_aperture} presented earlier, the specificity of these results is that they show an extinction phenomenon due to the absorption and scattering effect of the underwater channel with increased~$d_{\text{water}}$.

\subsection{Effect of Rx Parameters}
\begin{figure}
    \centering
    \includegraphics[width=0.485\textwidth]{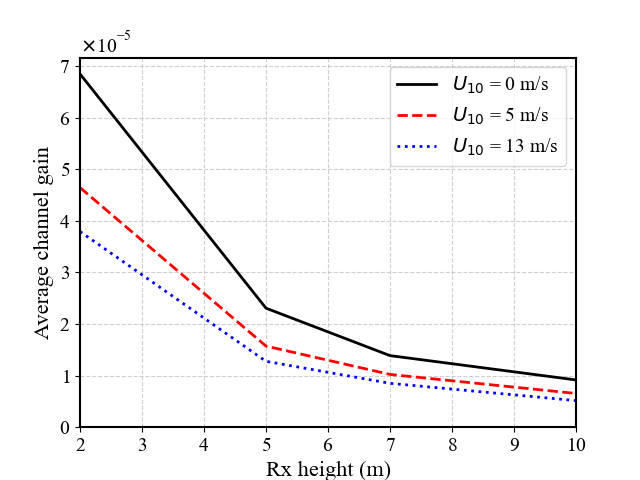}
    \caption{\small Average channel gain as a function of the \gls{rx} height $d_{\mathrm{air}}$ above the sea surface for different wind speeds $U_{10}$.}
    \label{Fig:gain_heights}
\end{figure}

First, let's study the effect of the \gls{rx} height relative to the sea surface, denoted by $d_{\text{air}}$ in Fig.\ref{fig:model_illustration}. 
We have shown in Fig.\,\ref{Fig:gain_heights} the average channel gain as a function of $d_{\text{air}}$, varying between $2$ and $10~\text{m}$.
As expected, increasing $d_{\text{air}}$ induces more channel loss. As previously established, the effects of scattering and absorption in air are neglected in this study. Thus, the observed losses result from the cumulative effect of photon refraction at the surface, the effect of \gls{uav} instability, combined with the increasing propagation distance in air, which drastically reduces the probability of photon reception on the \gls{pd}'s active area. The significantly higher average channel gain for the case of flat sea surface (i.e., $U_{10}=0$) further confirms this observation.

\begin{figure}
    \centering
    \includegraphics[width=0.485\textwidth]{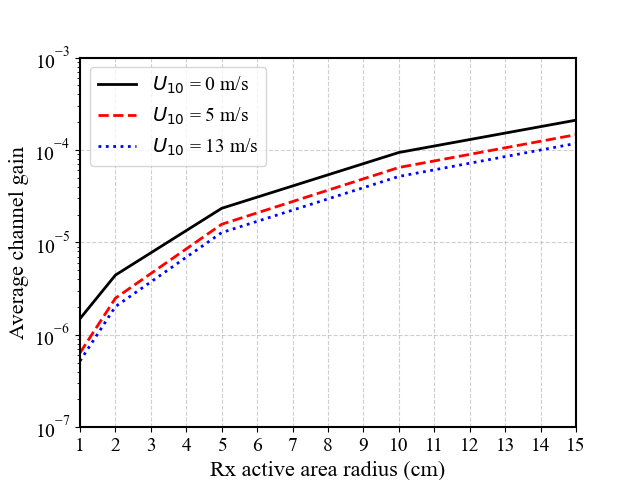}
    \caption{\small Average channel gain as a function of the \gls{rx} radius for different wind speeds $U_{10}$.}
    \label{Fig:gain_radius}
\end{figure}

We have further shown in Fig.\,\ref{Fig:gain_radius} the average channel gain as a function of the radius of the \gls{rx} equivalent active area. As expected, enlarging the \gls{rx} active area directly contributes to collecting a larger number of photons, and hence, to a larger channel gain. Note, in practice, the \gls{rx} active area can be dynamically adjusted alongside the \gls{rx} height using a mechanical iris mounted above the \gls{pd}, as proposed for example in \cite{Eltokhey-WCOM-2022}.
\subsection{Effect of Beam Misalignment}\label{Subsec:misalignement}
\begin{figure}
    \centering
    \includegraphics[width=0.25\textwidth]{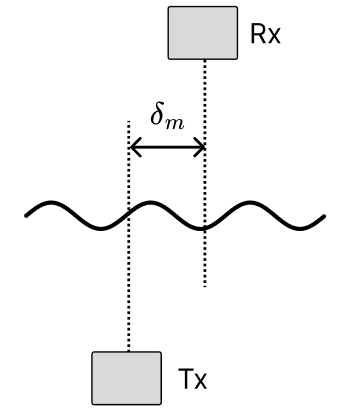}
    \caption{\small Illustration of the Tx-Rx horizontal displacement error (spatial offset) $\delta_m$.}
    \label{Fig:diagram_offset}
\end{figure}

In practice, the communication link is affected by \gls{tx}-\gls{rx} misalignment due mainly to the limited positioning accuracy of the underwater node or the \gls{uav}. In this paper, this effect is modeled by an offset horizontal displacement $\delta_m$ between the \gls{tx} and the \gls{rx} as illustrated in Fig.\,\ref{Fig:diagram_offset}. 

\begin{figure}
    \centering
    \includegraphics[width=0.485\textwidth]{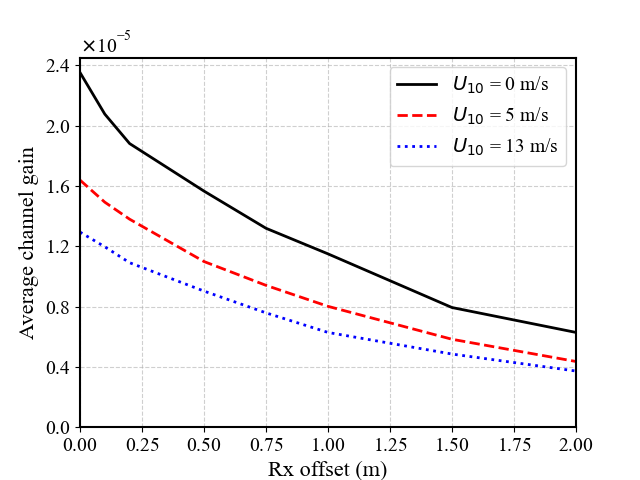}
    \caption{Average channel gain as a function of the \gls{tx}-\gls{rx} displacement $\delta_m$ for different wind speeds $U_{10}$.}
    \label{Fig:gain_offset}
\end{figure}

The effect of this spatial offset on the average channel gain is shown in Fig.\,\ref{Fig:gain_offset}.
We notice a certain robustness to \gls{tx}-\gls{rx} displacements thanks to the relatively large \gls{led} beam divergence angle ($\theta_{1/2}=10^\circ$). For instance, for $\delta_m = 0.5~\text{m}$, the corresponding loss with respect to $\delta_m=0$ is about $1.35~\text{dB}$ and $2.05~\text{dB}$ for $U_{10}=5$ and $13~\text{m/s}$, respectively, which is consistent with the results of Figs.\,\ref{fig:cov_aperture_10_ws_5} and \ref{fig:cov_aperture_10_ws_13}.

\subsection{E2E Link Performance Study}

Finally, this section presents the numerical results to evaluate the \gls{e2e} link performance in terms of the average \gls{ber} for a given data rate $R_b$ measured in \gls{mbps}.
The considered \gls{sipm} is the ONSE-30020 SiPM \cite{onsemi-SiPM} component with parameters summarized in Table\,\ref{tab:sipm_parameters} (see \cite{Khalighi-PhJ-2017} for additional description of the \gls{sipm} parameters). The considered \gls{tia} load resistance is $1~\text{k}\Omega$.

\begin{table}
    \centering
    \caption{ONSEMI MicroFJ-30020 SiPM parameters \cite{onsemi-SiPM}}
    
    \begin{tabular}{lcc}
        \hline
         Parameter & Symbol & Value \\
         \hline
         Gain & $G$ & $10^6$\\ 
         Number of \glspl{spad} & $N_\mathrm{SPAD}$ & $14410$\\ 
         Active area & $A_\mathrm{SiPM}$ & $9.43~\text{mm}^2$\\
         Fill factor & $F_F$ & $62~\%$ \\
         Photon detection efficiency & $\Upsilon_\mathrm{PDE}$ & $30~\%$ \\ 
         Dark count rate & $f_\mathrm{DCR}$ & $471~\text{kHz}$\\ 
         Cross-talk probability & $P_\mathrm{CT}$ & $2.5~\%$ \\ 
         After-pulsing probability  & $P_\mathrm{AP}$ & $0.75~\%$ \\ 
         \hline
    \end{tabular}
	\label{tab:sipm_parameters}
\end{table}

As for the \gls{tx}, we use the NICHIA NSPB510AS \gls{led} \cite{LED-spec} with central wavelength $\lambda = 470~\text{nm}$ and set the transmit power to $P_{\text{Tx}} = 100~\text{mW}$; this wavelength corresponds to the maximum sensitivity of the selected \gls{sipm}.
We consider the simple uncoded \gls{nrz} \gls{ook} modulation with extinction ratio $\xi_\mathrm{OOK} = 0.1$. \gls{nrz}-\gls{ook} has the advantage of implementation simplicity assuming the need to relatively low to moderate data rates.
Signal demodulation is performed based on optimal threshold calculation, see \cite{Ijeh-JOE-2021} for details. Additionally, we take into account the modulation bandwidth limitation of the \gls{sipm} and the \gls{led}, which are experimentally evaluated as $f_{\text{c,LED}} = 10~\text{MHz}$ and $f_{\text{c,SiPM}} = 2~\text{MHz}$, and model their frequency responses as first-order low-pass filters, see \cite{Khalighi-PhJ-2017, Khalighi-JOE-2019}.

\begin{figure}
    \centering
    \includegraphics[width=0.485\textwidth]{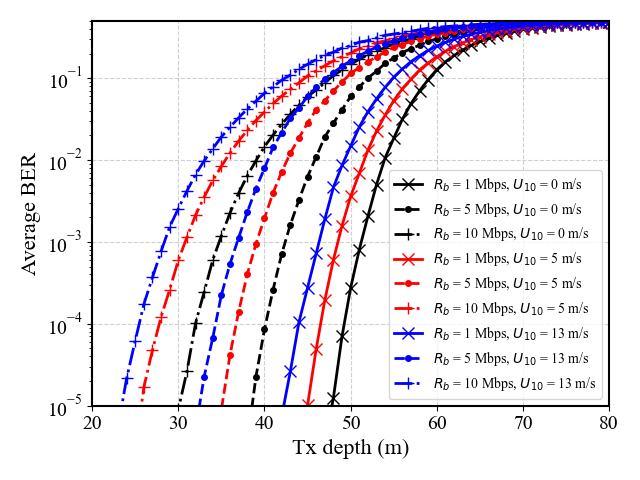}
    \caption{\small Average BER as a function of Tx depth for bit rates $R_b=1, \,5, \,10$\,Mbps and wind speeds $U_{10}=0,\, 5,\,13$\,m$/$s.}
    \label{Fig:BER_z}
\end{figure}

Figure\,\ref{Fig:BER_z} shows \gls{ber} plots as a function of the \gls{tx} depth $d_{\text{water}}$ for the wind speeds of $U_{10}=5$ and $13~\text{m/s}$, and different data rates $R_b$, assuming $d_{\text{air}}=5~\text{m}$. As benchmark, we have also shown the \gls{ber} plots for \mbox{$U_{10}=0$}. 

First, we notice the performance degradation with increase in $R_b$, which is due to the limited bandwidth of the \gls{sipm} and the \gls{led}, resulting in reduced link range \cite{Khalighi-PhJ-2017}. For instance, for a target \gls{ber} of $10^{-3}$ and $U_{10}=0$\,m/s, the maximum \gls{tx} depth $d_{\text{water}}$ for $R_b = 5$ and $10~\text{Mbps}$ is about $43$ and $36~\text{m}$, respectively. Note that, frequency-domain equalization can be employed to improve the effective link range, as considered in \cite{Khalighi-JOE-2019}.
Second, we notice the substantial performance degradation with the surface wind effect. For instance, for $U_{10}=5~\text{m/s}$ and the same target \gls{ber} of $10^{-3}$, the link range is reduced by $4$ and $5$\,m for $R_b = 5$ and $10~\text{Mbps}$, respectively, with respect to the flat surface case (i.e., $U_{10}=0$). We can also observe the performance degradation with increase in $U_{10}$: for a target \gls{ber} of $10^{-3}$, the link range is reduced by $\approx2.5~\text{m}$ by increasing $U_{10}$ from $5$ to $13~\text{m/s}$. 

Lastly, considering a typical data rate of $R_b = 1~\text{Mbps}$ in the \gls{iout} context for transmitting underwater sensor data, and given the \gls{fec} limit of $2\times 10^{-3}$ for the \gls{ber}, it is quite feasible to establish \gls{uwoc} links with \gls{tx} depths up to $\sim47$\,m with surface wave speeds reaching up to $13~\text{m/s}$. This range of depth is in fact sufficient for the intended application of coral reef monitoring. 
It is worth noting that the obtained performance is mainly due to the use of a highly sensitive \gls{sipm} as the \gls{pd}, and to the fact that the transmission is assumed to take place at night, i.e., in the absence of background solar noise \cite{Hamza-OpEX-2016, Hamza-WACOWC-2019}.

To further validate the feasibility of reliable link deployment, we evaluate the \gls{nep} of the \gls{sipm} using the following expression \cite{NEP-SiPM}:
\begin{equation}
    \mathrm{NEP}(\lambda) = E_\mathrm{ph} \frac{F_e \sqrt{f_{\text{c,SiPM}}}}{\Upsilon_\mathrm{PDE}} 
    \left( 1 + \sqrt{1 + \frac{2 f_\mathrm{DCR}}{f_{\text{c,SiPM}}}} \right),
    \label{eq:NEP}
\end{equation}
where $E_\mathrm{ph}=h_p c_p / \lambda$ denotes the photon energy corresponding to the wavelength $\lambda$, with $h_p$ and $c_p$ being Planck’s constant and the light celerity, respectively. Also, $\Upsilon_\mathrm{PDE}$ and $f_\mathrm{DCR}$ represent the \gls{sipm}'s photon detection efficiency and dark count rate, respectively, and $F_e$ is its excess noise factor, given by \cite{NEP-SiPM}:
\begin{equation}
    F_e = \frac{1}{1 + \ln(1 - P_\mathrm{CT} - P_\mathrm{AP})}.
    \label{eq:F}
\end{equation}
Here, $P_\mathrm{CT}$ and $P_\mathrm{AP}$ denote the probabilities of optical cross-talk and after-pulsing, respectively. The \gls{sipm}'s \gls{nep}, in fact, represents the incident optical power required to yield a unit \gls{snr} at the detector output. Accordingly, the equivalent optical noise power can be expressed as $P_n = \mathrm{NEP}(\lambda)\sqrt{B}$, where $B$ denotes the effective noise bandwidth \cite{Mackowiak-Thorlabs-2015}.

For the selected \gls{sipm} with parameters listed in Table~\ref{tab:sipm_parameters}, we obtain a \gls{nep} of $3.42\times10^{-15}~\mathrm{W/\sqrt{Hz}}$ at the considered wavelength $\lambda=470$\,nm. Assuming the approximate equivalent noise bandwidth as $B \approx R_b/2$ for the considered \gls{nrz}-\gls{ook} signaling, the corresponding equivalent noise power equals $P_n = 2.42\times 10^{-12}$, \mbox{$5.41\times 10^{-12}$}, and $7.65\times 10^{-12}$\,W for $R_b=1,\,5,$ and $10$~\gls{mbps}, respectively \cite{Mackowiak-Thorlabs-2015}. This ensures a sufficiently high \gls{snr} for reliable communication over the considered ranges. For instance, for $U_{10}=13$~m/s and $R_b=1$~\gls{mbps}, the system achieves an average electrical \gls{snr} of approximately $15.63$~dB for $d_\mathrm{water}=47$\,m at the \gls{sipm} output, resulting in a \gls{ber} below the \gls{fec} limit (as mentioned in the last paragraph of the previous page). Neglecting the effects of surface waves, air bubbles, and UAV instability, using \cite[Eq.~3]{Ijeh-JOE-2021}, the corresponding \gls{snr} is about $28.64$~dB. This means we need a link margin of about $13$\,dB to compensate the channel random effects in this example.


\section{Conclusions}
\label{sec:concluson}
In this work, we presented a comprehensive channel characterization and modeling for \gls{w2a} optical wireless communications for the purpose of marine ecosystem monitoring.
The proposed approach uses a Monte Carlo-based ray-tracing algorithm that incorporates key parameters of the underwater channel on photon propagation, and accounts for the effects of sea surface through the \gls{jonswap} spectrum model, as well as air bubbles near the sea surface through Mie theory and the \gls{hn} model. In addition, we proposed a new model to characterize the impact of \gls{uav} instability under turbulent wind conditions on the \gls{w2a} link. This way, critical environmental factors such as wind speed and distance to the seashore, particularly relevant for coral reef studies, are taken into account.

Based on the thoughtful choice of \gls{jonswap} and \gls{hn} models and considering realistic parameters for a typical \gls{iout} system for coral reef monitoring, we investigated the impact of the instability of the \gls{uav}, the wind speed and air bubbles, \gls{tx} and \gls{rx} parameters, including as well the impact of link misalignment. The \gls{e2e} link performance analysis demonstrated the viability of the proposed system for coral reef monitoring, achieving a data rate of $1$\,\gls{mbps} at depths of up to $47$ meters and with wind speeds of up to $13$\,m/s. 

The relative dominance of the considered link parameters depends strongly on the operating conditions. While the overall link budget is primarily constrained by underwater depth (due to severe absorption and scattering), the dynamic sea surface emerges as the main source of signal degradation, inducing significant beam refraction that becomes more pronounced at higher \gls{uav} altitudes. In contrast, signal attenuation caused by \gls{uav} tilt instability under turbulent wind is comparatively marginal.
This is largely due to the use of a wide-beam \gls{led}: although its larger divergence increases susceptibility to surface refraction losses, it provides strong robustness against misalignment, effectively mitigating pointing errors, which are a common challenge in locating the underwater \gls{tx}.

In summary, this work provides essential insights and guidelines for designing \gls{iout} systems with \gls{w2a} optical links tailored to marine biodiversity monitoring. The proposed photon propagation algorithm serves as a valuable tool for further exploration of the \gls{w2a} communication channel, in view of developing robust systems capable of mitigating the effects of bubbles and sea surface dynamics. 
{\appendices
\section{Temperature and Salinity Profiles in Coral Reef Waters}\label{app:temp_sal}

Oceanic turbulence has been extensively investigated in the field of \gls{uwoc}, primarily in the context of vertical links \cite{Ijeh-JOCN-2022, Elamassie-TCOM-2020}. This is because the phenomenon is typically caused by temperature and salinity gradients across different ocean layers, which can lead to deviations of the optical beam \cite{OceanOptics_Website}. In fact, these gradients are not uniform across all oceans, as illustrated in \cite[Fig. 1]{Elamassie-TCOM-2020}. As briefly discussed in Section~\ref{sec:System_Model}, in this work, we assume that the temperature and salinity gradients are negligible, which allows us to neglect the effects of oceanic turbulence in our \gls{w2a} channel modeling. To support this assumption, we present in this Appendix the temperature and salinity gradients observed in reef waters near the Great Barrier Reef of New Caledonia.

\begin{figure}
    \centering
    \includegraphics[width=0.485\textwidth]{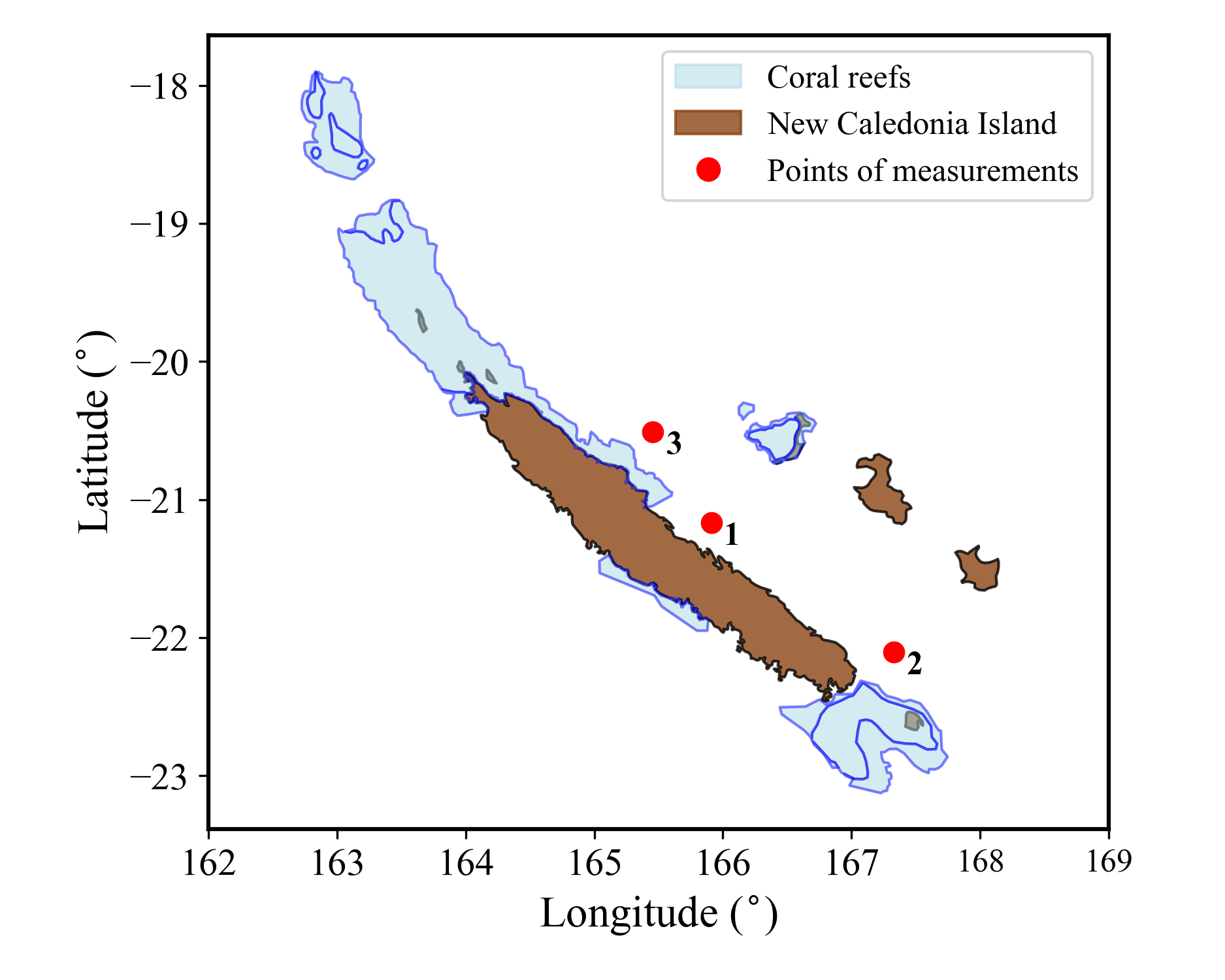}
    \caption{\small Geographical location of the three AODN temperature and salinity measurement stations studied in this work based on data from \cite{AODN-Portal}.}
    \label{Fig:coral_reef_map}
\end{figure}
Our analysis relies on measurement data provided by the \gls{aodn} \cite{AODN-Portal}. As shown in Fig.\,\ref{Fig:coral_reef_map}, we consider three measurement stations located near the reefs of New Caledonia, at coordinates $(165.91^\circ\text{ E},\, 21.17^\circ\text{ S})$, $(167.33^\circ\text{ E},\, 22.11^\circ\text{ S})$, and $(165.45^\circ\text{ E},\, 20.51^\circ\text{ S})$, respectively. We have shown in Figs.\,\ref{temp_profile} and \ref{sal_profile} the vertical profiles of temperature and salinity, respectively, as functions of ocean depth. It can be clearly observed that within the first 50 meters of depth (the region of interest in this study), both temperature and salinity gradients are extremely weak. These results, corroborated by MacKellar \textit{et al.} \cite{Mackellar-JGRA-2013}, indicate that oceanic turbulence effects can be considered negligible in the context of our analysis.


\begin{figure}
    \centering
    \subfloat[]{\includegraphics[width=0.485\textwidth]{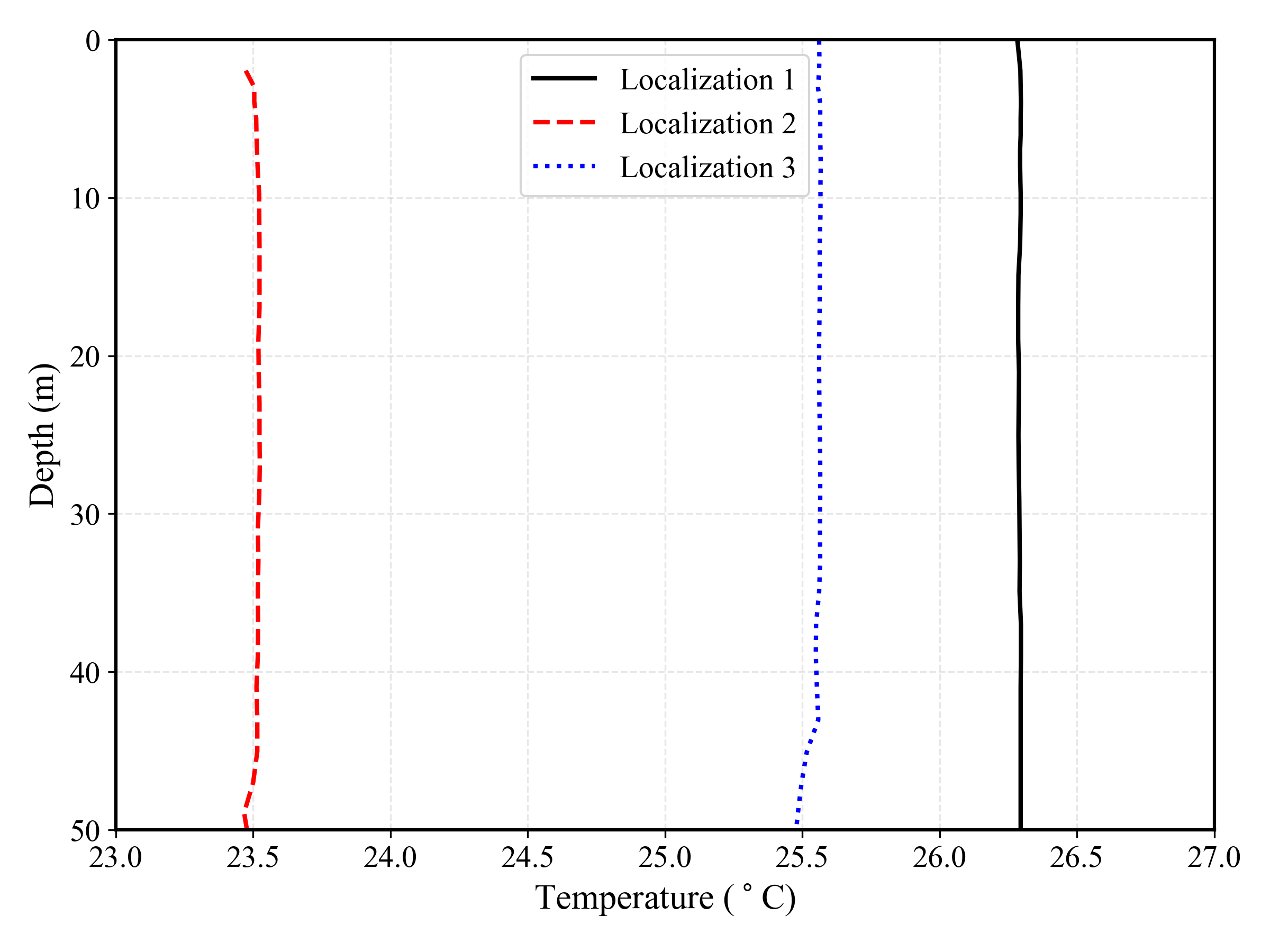}%
    \label{temp_profile}}
    \newline
    \subfloat[]{\includegraphics[width=0.485\textwidth]{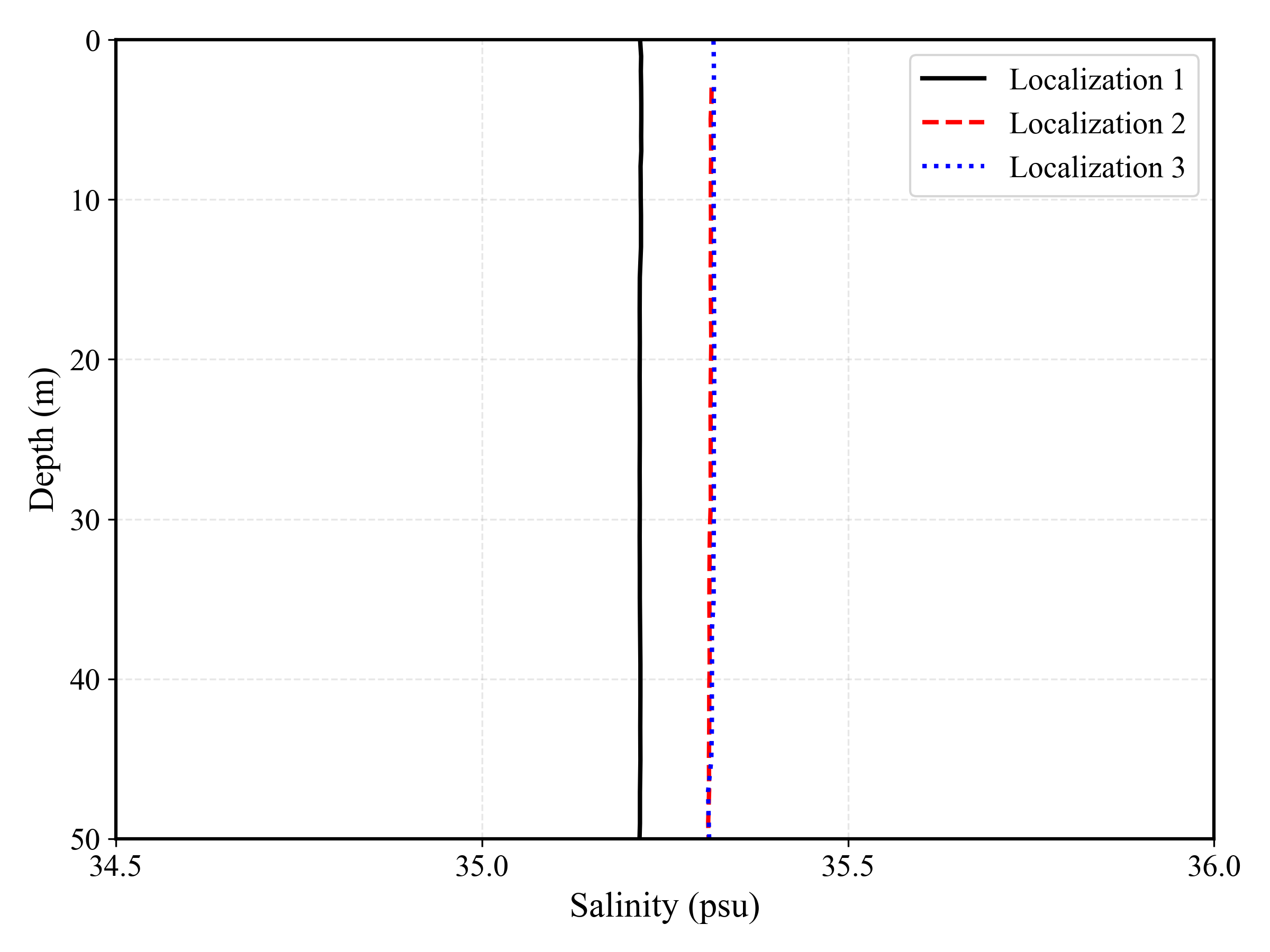}%
    \label{sal_profile}}
    \caption{\small Oceanic profiles of (a) temperature (b) salinity in three localizations near the New Caledonia Barrier reef indicated in Fig.\,\ref{Fig:coral_reef_map}. psu denotes practical salinity unit. Results are based on the data provided in \cite{AODN-Portal}.}
    \label{temp_sal}
\end{figure}

\section{Turbulent Wind Modeling According to Dryden Model}\label{app:Dryden_implementation}

We consider in this work turbulent winds (i.e., with time-varying wind speed), which are commonly encountered in analyses of airflow over the ocean. To model the effect of these phenomena, here we employ the Dryden model \cite{Dryden-1990}, which is one of the most widely adopted approaches in the literature \cite{Wang-MeasCont-2019, Kachroo-TAP-2021}. This model is derived from extensive measurements and experimental data, providing an empirical characterization of the \gls{psd} of wind turbulence.

Following the methodology described in \cite{Kachroo-TAP-2021}, the wind is modeled as the combination of a average wind component $U=(u_x, u_y, u_z)^\top$ and a turbulent component $\widetilde{U} = (\tilde{u}_x, \tilde{u}_y, \tilde{u}_z)^\top$. The Dryden model generates the turbulence velocities by filtering three independent, unit variance, band-limited white Gaussian noise processes, one for each spatial direction, through transfer functions defined as follows \cite{Kachroo-TAP-2021}:
\begin{align}
    \label{eq:dryden_transfer_equation}
    \begin{aligned}
        H_x(s) & =\sigma_x \sqrt{\frac{2 L_x}{\pi |U|}} \frac{1}{\left(1+\frac{L_x}{|U|}s\right)}, \\
        H_y(s) & =\sigma_y \sqrt{\frac{L_y}{\pi |U|}} \frac{\left(1+\frac{\sqrt{3}L_y}{|U|}s\right)}{\left(1+\frac{L_y}{|U|}s\right)^2}, \\
        H_z(s) & =\sigma_z \sqrt{\frac{L_z}{\pi|U|}} \frac{\left(1+\frac{\sqrt{3}L_z}{|U|}s\right)}{\left(1+\frac{L_z}{|U|}s\right)^2},
    \end{aligned}
\end{align}
where $\sigma_x$, $\sigma_y$, and $\sigma_z$ denote the wind turbulence standard deviations, and $L_u$, $L_v$, and $L_w$ represent the turbulence length scales in directions $x$, $y$, and $z$, respectively.

In this study, we follow the military standard MIL-F-8785C \cite{Military_report}, which provides the expressions for turbulence length scales as follows: 
\begin{align}
    \label{eq:dryden_length_scale}
    \begin{aligned}
        L_z & = d_{air},\\
        L_x & = L_y = \frac{d_{air}}{(0.177 + 0.000823d_{air})^{1.2}}.
    \end{aligned}
\end{align}
At low altitudes (below $<1000$ feet), the standard deviations take the following values \cite{Military_report}:
\begin{align}
    \label{eq:dryden_std}
    \begin{aligned}
        \sigma_z & = 0.1\,W_{20},\\
        \frac{\sigma_x}{\sigma_z} & = \frac{\sigma_y}{\sigma_z} = \frac{1}{(0.177 + 0.000823\,d_{air})^{0.4}},
    \end{aligned}
\end{align}
where $W_{20}$ denotes the wind speed at $20$ feet ($\approx 6$ m) in knots. It is typically set to $15$\,knots for low turbulence, $30$\,knots for moderate turbulence, and $45$\,knots for high turbulence conditions.

Figures~\ref{wind_x} and \ref{wind_y} illustrate an example of variations of the simulated wind for $u_x=u_y=5$\,m/s along the $x$ and $y$ axes, respectively, under moderate turbulence conditions.

\begin{figure}
    \centering
    \subfloat[]{\includegraphics[width=0.485\textwidth]{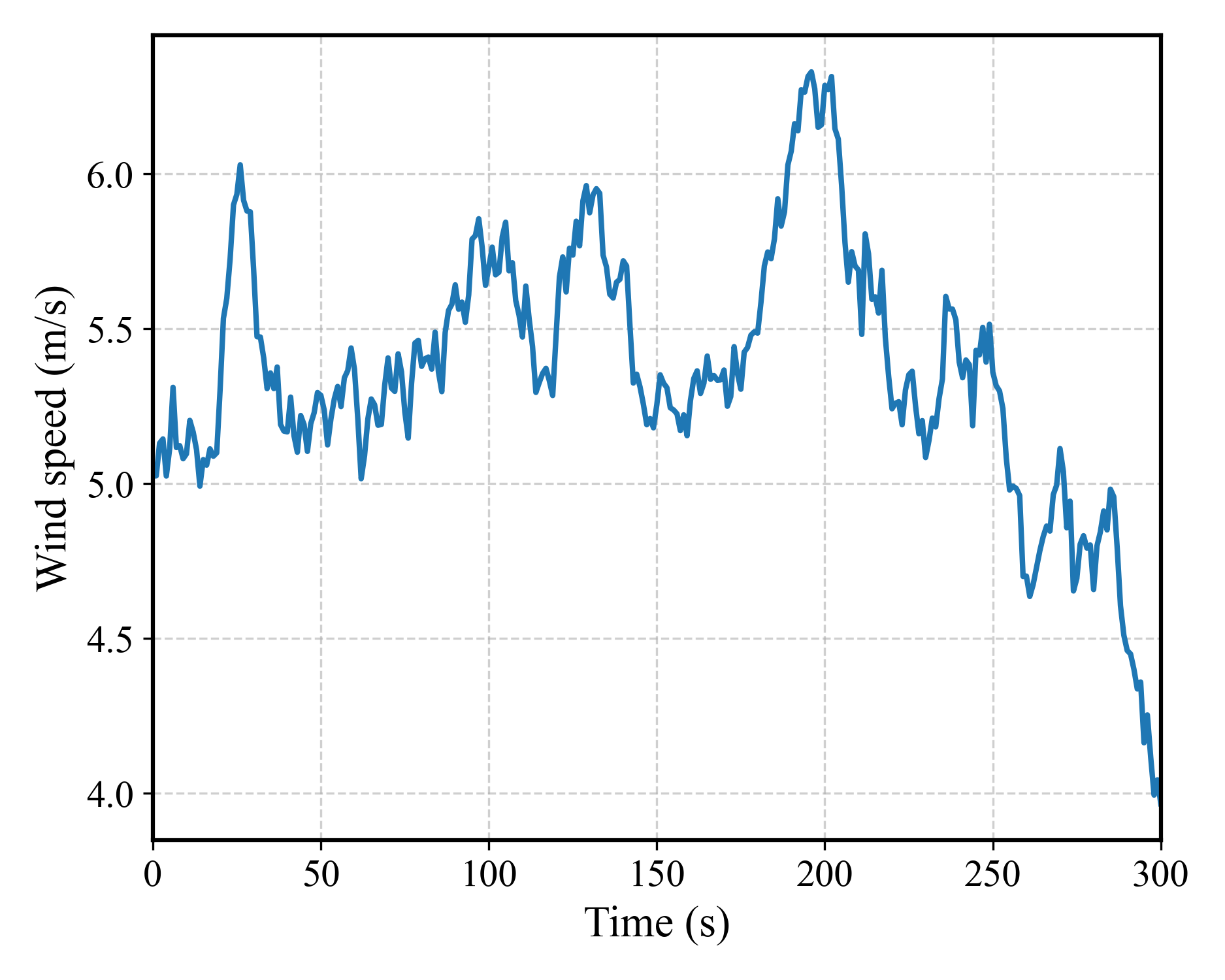}%
    \label{wind_x}}
    \newline
    \subfloat[]{\includegraphics[width=0.485\textwidth]{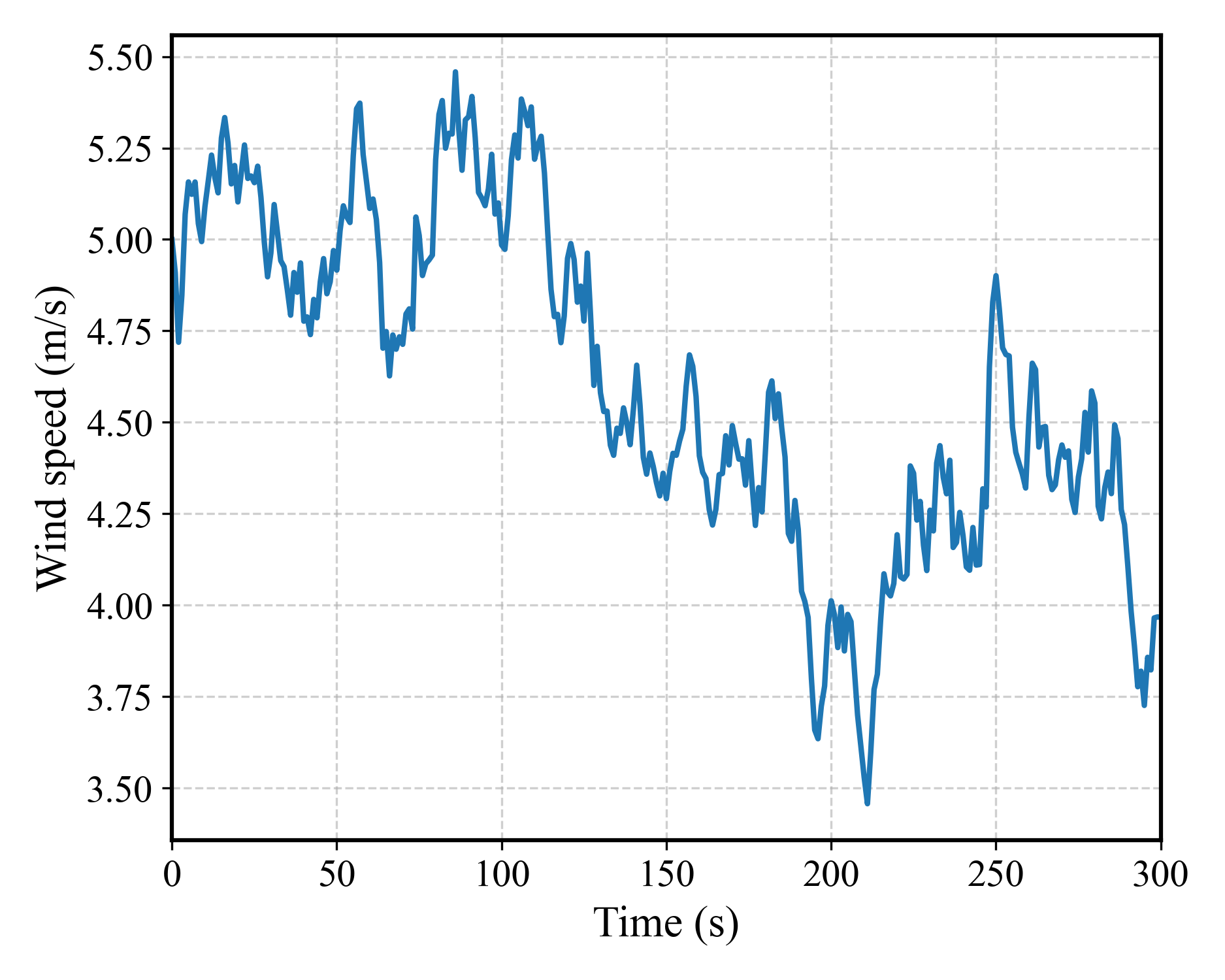}%
    \label{wind_y}}
    \caption{\small Variations of wind speed for average wind speed of $u_x=u_y=5$ m/s (under moderate turbulence conditions) along the (a) $x$-axis and (b) $y$-axis.}
    \label{wind_x_y}
\end{figure}

\section{Distributions of Turbulent Wind Components}\label{app:distribution_proof}

In this appendix, we derive the statistical distributions of the wind velocity components, i.e., $V = (V_x, V_y, V_z)$, using the Dryden model. To this end, we recall the linear filters $H_x$, $H_y$, and $H_z$ introduced in (\ref{eq:dryden_transfer_equation}). The inputs to these filters are zero-mean Gaussian white noise processes with unit variance, i.e., $\mathcal{N}(0,1)$. 
Consequently, the outputs of the filters can be also be modeled as Gaussian random processes where, for instance, the variance corresponding to $H_x$ output given by:
\begin{equation}
    \sigma^2_\mathrm{out,x} = \int_{-\infty}^{+\infty} |h_x(t)|^2 \, dt,
\end{equation}
where $h_x(t)$ denotes the filter impulse response.

For the variance of $V_x$, we recall the transfer function of $H_x$ as a function of the Laplace operator $s$:  
\begin{equation}
    H_x(s) = \sigma_x \sqrt{\frac{2 L_x}{\pi U}} \frac{1}{1 + \tfrac{L_x}{U}s} 
    = \frac{K}{1 + \tau s},
\end{equation}
where $K = \sigma_x \sqrt{\tfrac{2 L_x}{\pi U}}$ and $\tau = L_x / U$, with $U$ denoting the average wind speed and $L_x$ the turbulence length scale of $H_x$, respectively. The corresponding impulse response $h_x(t)$ is obtained using the inverse Laplace transform:  
\begin{equation}
    h_x(t) = K \cdot \frac{1}{\tau} e^{-t/\tau}\, u_H(t),
\end{equation}
where $u_H(t)$ is the Heaviside unit step function. Then, the output variance is obtained by evaluating the energy of the filter:  
\begin{equation}
    \sigma_{V_x}^2 = E_{h_x} 
    = \frac{K^2}{\tau^2}\int_0^{\infty} e^{-2t/\tau} dt 
    = \frac{K^2}{2\tau} 
    = \frac{\sigma_x^2}{\pi}.
\end{equation}

A similar procedure applies to $V_y$ and $V_z$, since $H_y$ and $H_z$ exhibit similar structures. For the sake of clarity, we provide details on the derivation for $H_y$ in the following:  
\begin{equation}
    H_y(s) = K \frac{1 + \sqrt{3}\tau s}{(1 + \tau s)^2},
\end{equation}
with $K = \sigma_y \sqrt{(2 L_y)/(\pi U)}$ and $\tau = L_y/U$. By applying the inverse Laplace transform, we obtain:
\begin{equation}
    h_y(t) = K \left( \frac{\sqrt{3}}{\tau} + \frac{1 - \sqrt{3}}{\tau^2}t \right) e^{-t/\tau} u_H(t).
\end{equation}
Exploiting the following general formula,
\begin{equation}
    \int_0^{\infty} t^n e^{-2t/\tau} dt = \frac{n!}{(2/\tau)^{n+1}},
\end{equation}
the total energy of the filter simplifies to:
\begin{align}
    \begin{aligned}
        E_{h_y}  & = \int_0^\infty h_y(t)^2 dt \\
    & = K^2 \int_0^\infty \left( \frac{\sqrt{3}}{\tau} + \frac{1 - \sqrt{3}}{\tau^2}t \right)^2 e^{-2t/\tau} dt \\
    & = \frac{K^2}{\tau} 
    = \frac{\sigma_y^2}{\pi}.
    \end{aligned}
\end{align}By analogy, the same result holds for $V_z$. Thus, we establish that:
\begin{equation}
    \sigma_{V_j}^2 = \frac{\sigma_j^2}{\pi}, \quad j \in \{x,y,z\}.
\end{equation}
Finally, by combining the turbulent components with the constant mean wind $U$, the resulting velocity components follow a Gaussian distribution as follows:  
\begin{equation}
    V_j \sim \mathcal{N}(u_j, \, \sigma_j^2/\pi), \quad j \in \{x,y,z\}.
\end{equation}}

\section*{Disclosures} 
The authors declare no conflicts of interest. 

\section*{Acknowledgment}
The authors would like to thank Mr. Yves Chardard from SubSeaTech Co., Marseille, France, and Mr. Djamel Eddine Tahraoui from Segula Technologies Co., for their valuable guidance and insightful advice regarding the modeling aspects of the \gls{uav}. They also extend their sincere gratitude to Dr. Mehmet Cagri Ilter, from Nokia, Finland, for his valuable feedback and constructive remarks. 

\balance 
\bibliography{IEEEabrv,biblio}{}
\bibliographystyle{IEEEtran}

\vfill

\end{document}